\documentclass[11pt,a4paper]{article}
\pdfoutput=1

\usepackage{chet}
\usepackage{amsmath,amssymb,amsthm}
\usepackage{xcolor}
\usepackage{bm}
\usepackage{graphicx}
\usepackage{slashed}

\hypersetup{colorlinks,citecolor=blue}
\newcommand{\dd}{\text{d}}

\newcommand{\vev}[1]{\langle #1 \rangle}
\newcommand{\tr}{\operatorname{tr}}

\title{Anomalies in Neural Network Field Theory}

\author{Christian Ferko$^{\ast,\dagger,}$\email{c.ferko@northeastern.edu},
Samuel Frank$^{\ast,}$\email{frank.sam@northeastern.edu},
James Halverson$^{\ast,\dagger,}$\email{j.halverson@northeastern.edu}, and
Vishnu Jejjala$^{\ddagger,\ast,\dagger,}$\email{v.jejjala@wits.ac.za}}

\affiliation{$^{\ast}$Department of Physics, Northeastern University, Boston, MA 02115, USA\\
$^{\dagger}$The NSF Institute for Artificial Intelligence and Fundamental Interactions (IAIFI)\\
$^{\ddagger}$Mandelstam Institute for Theoretical Physics, School of Physics, and NITheCS,\\
University of the Witwatersrand, Johannesburg, WITS 2050, South Africa}

\abstract{

Neural network field theory (NN-FT) formulates field theory in terms of a network architecture and a density on its parameters.
We derive Schwinger--Dyson equations and Ward identities in NN-FT and utilize them to study anomalies.
The equations depend on a conserved parameter space current that characterizes symmetries and how they break. It is relevant even in non-local NN-FTs, but can recover local currents in the case of a local Lagrangian by an appropriate fiber-wise average.
In machine learning, this formalism is applied to feedforward networks and the attention mechanism.
In physics, we use this machinery to study $U(1)$ symmetry for a complex scalar, the scale anomaly in $4d$ massless $\phi^4$ theory, the Weyl anomaly for the bosonic string (including a new computation of the critical dimension), and examples involving discrete topological data, such as winding numbers and T-duality.
Since the results are obtained in network parameter space rather than the standard field space, they represent a new way to understand symmetries in quantum field theories. 
}

\begin{document}
\maketitle

\toc

\clearpage
\section{Introduction}
\label{sec:intro}

In classical field theory, Noether's theorem states that every continuous symmetry of the action gives rise to a conserved current satisfying $\partial_\mu j^\mu = 0$ on-shell. In quantum field theory, however, the path integral sums over all field configurations, including off-shell fluctuations. The quantum analogues of the classical equations of motion are the Schwinger--Dyson equations, which show that operator insertions of the classical equations of motion vanish inside correlation functions up to contact terms proportional to Dirac delta functions.
Ward identities arise from applying symmetry transformations as changes of variables in the path integral. They implement Noether's theorem quantum mechanically, showing that current conservation holds inside correlation functions up to contact terms that encode the symmetry transformations of operator insertions. Thus, while $\partial_\mu j^\mu$ does not vanish identically inside Green's functions, its insertions enforce the symmetry structure of the fully quantized theory.
An anomaly occurs when a symmetry of the classical action fails to preserve the quantum path integral measure $\mathcal{D}\phi$. Under a symmetry transformation, one may have $\mathcal{D}\phi' = J \mathcal{D}\phi$ with a non-trivial Jacobian $J \neq 1$. Fujikawa~\cite{Fujikawa1979} showed that this Jacobian directly produces the anomalous divergence of the Noether current. A classic example is the Adler--Bell--Jackiw anomaly responsible for $\pi^0 \to \gamma\gamma$~\cite{Adler1969,BellJackiw1969}, a physical manifestation of a global anomaly. Gauge anomalies, on the other hand, are especially severe because they obstruct consistent quantization.

Many Euclidean QFTs on $\mathbb{R}^d$ can be expressed in terms of neural network architectures and parameter densities, as first proposed in~\cite{Halverson:2020trp,Halverson:2021aot}.
A random draw of the parameters $\theta$ with respect to an appropriately chosen distribution determines a field configuration\footnote{Strictly speaking, typical field configurations in QFT are Schwartz distributions rather than functions, but this subtlety can be addressed in NN-FT~\cite{Ferko:2026axm} so we will be cavalier the distinction in what follows.} $\phi_\theta:\mathbb{R}^d\to\mathbb{R}$, and correlation functions are computed as ordinary expectations over parameter space,
\begin{align}
\langle \mathcal{O} \rangle &:= \mathbb{E}[\mathcal{O}] = \int d\mu(\theta)\,{\cal O}[\phi_\theta] \,, \label{eq:vev} \\
G_n(x_1,\ldots,x_n) &= \int d\mu(\theta) \prod_{i=1}^n \phi_\theta(x_i) \,. \label{eq:corr}
\end{align}
Here, $d\mu(\theta) = d\theta\, p(\theta)$.
The functional measure is not postulated directly; it is induced by pushing forward the probability measure on network parameters.
This viewpoint turns architecture, activation functions, and parameter densities into the microscopic data from which the Schwinger functions of a field theory are built.

A particularly important mechanism in the correspondence is the large width limit of a single hidden layer neural network.
For example, in a random feature model
\begin{equation}
\phi_N(x)=\frac{1}{\sqrt N}\sum_{a=1}^N A_a\,\sigma(W_a\cdot x+B_a) \,,
\end{equation}
the values of $\phi_N$ at finitely many spacetime points become jointly Gaussian as $N\to\infty$, under standard parameter and moment assumptions~\cite{neal}.
This is essentially a consequence of the central limit theorem.
The resulting Gaussian process is a generalized free field whose covariance is fixed by the architecture and the parameter distribution.
This is a constructive realization of free field theory that applies to many quantum systems in the infinite width limit of the neural network. Interactions arise from controlled departures from this limit: finite width produces non-vanishing higher cumulants, while correlated or non-Gaussian parameter densities can generate connected correlators even at infinite width.
Under suitable analyticity, positivity, and locality assumptions, the connected correlators can be used to reconstruct an effective action, often perturbatively or order by order in an expansion parameter.

The emerging picture is that NN-FT is both a new theoretical framework for QFT and a new method for simulating \emph{continuum} QFTs on a computer. There has been considerable recent progress on both fronts. In addition to realizing bosonic theories such as $\phi^4$~\cite{Demirtas:2023fir}, NN-FTs can be used to represent fermions and supersymmetric theories~\cite{Huang:2025ipy,Frank:2025zuk}; conformal field theories~\cite{Halverson:2024axc}, like the Liouville CFT~\cite{Ferko:2026axm}, and their Virasoro symmetry~\cite{Robinson:2025ybg}, along with conformal defects~\cite{Capuozzo:2025ozt}; worldsheet string theory~\cite{Frank:2026bui,Ageev:2026ofv}; and topological effects in QFT~\cite{Ferko:2026ken}. In~\cite{Ferko:2026axm} it was shown that any QFT associated with a positive measure over field configurations admits a NN-FT description, generalizing an analogous result in quantum mechanics~\cite{Ferko:2025ogz}, so this framework is fairly universal.\footnote{Although this universality theorem is non-constructive, and the architecture and parameter density whose existence is guaranteed by the theorem are not known to be ``optimal'' in any sense, there are optimality results for other classes of architectures such as random Fourier features~\cite{Zhang:2026tss}.} See also~\cite{Hashimoto:2024aga,Sen:2025vzl,Erbin:2021kqf} for an incomplete sampling of other NN-FT results and~\cite{Halverson:2024hax} for an introduction to related machine learning ideas for physicists.

In NN-FT, symmetries can be imposed in the space where the theory is defined.
When $g$ is a symmetry operation, it acts on the neural network architecture in a way that can often be absorbed into the parameters
\begin{equation}
g:\ \theta \;\mapsto\; \theta'(\theta, g)
\end{equation}
preserving the structure of~\eqref{eq:vev} and~\eqref{eq:corr}~\cite{Halverson:2021aot}, demonstrating the existence of a symmetry. This absorption mechanism is the source of most work on symmetries in NN-FT to date.

In this paper we push symmetries in NN-FT further, developing the analogues of Schwinger--Dyson equations, Ward identities, and anomalies directly in neural network parameter space.
For a parameter density $p(\theta)$ and any smooth observable $\mathcal{O}$, the vanishing of a total derivative in a parameter direction gives
\begin{equation}
\bigl\langle \partial_a\mathcal{O}\bigr\rangle = -\bigl\langle (\partial_a\log p)\,\mathcal{O}\bigr\rangle \,.
\end{equation}
This identity is exact at finite width and does not assume a continuum limit, locality, Gaussianity, or a special architecture.
Contracting it with an arbitrary vector field $\xi=\xi^a\partial_a$ on parameter space gives a general identity
\begin{equation}
\bigl\langle \delta_\xi\mathcal{O}\bigr\rangle = -\bigl\langle B_\xi(\theta)\,\mathcal{O}\bigr\rangle \,,\qquad
B_\xi(\theta)=\sum_a \xi^a\,\partial_a\log p+\sum_a \partial_a\xi^a \,,
\end{equation}
up to boundary flux terms when the parameter domain has a boundary.
It is convenient to call $\partial_a\log p$ the \emph{score}, $s_a$.
We dub $B_\xi(\theta)$ the \emph{breaking function}, which is closely related to a parameter space current, discussed below.
It is not defined only for symmetries: every flow in parameter space has an associated breaking function.

Ward identities arise when the parameter space flow absorbs a transformation of the output fields.
In this case, $\delta_\xi\phi_\theta$ can be written in terms of the symmetry action on $\phi_\theta$, so the left hand side of the contracted Schwinger--Dyson equation becomes the usual variation of field insertions.
The right hand side measures the failure of the parameter space measure to be invariant.
This failure separates into two pieces.
The sum $\sum_a\xi^a\,s_a$ is the breaking due to the score of the chosen parameter density, while $\sum_a\partial_a\xi^a$ is the Jacobian of the flat parameter measure.
In continuum limits, a regulator dependent remnant of such Jacobians can become the analogue of a Fujikawa anomaly.
When boundary fluxes are absent, $B_\xi(\theta)=0$ is equivalent to the continuity equation $\sum_a\partial_a(\xi^a p)=0$ and gives an unbroken Ward identity.
When boundaries or compact cutoffs are present, the same total derivative computation keeps track of the compensating flux explicitly.

The second purpose of this work is to relate the parameter space picture --- which is universal in the scalar sector and can describe non-local theories ---  to the more familiar field space story, including theories defined by local Lagrangians.
The network map pushes the parameter density forward to a distribution over field configurations and hence to an effective field space action.
For observables that depend on $\theta$ only through $\phi_\theta$, expectation values computed in parameter space and in field space agree.
Consequently, the field space breaking function is the density weighted fiber average of $B_\xi(\theta)$ over all parameter configurations that realize the same field.
If the induced field theory is local, this field space breaking function can be written as the integrated divergence of a Noether current plus the usual field space anomaly.
The parameter space current $j^a=\xi^a p$ and the spacetime Noether current are not the same object, but they are connected by pushforward, fiber averaging, and locality.

The organization of this paper is as follows: Sections~\ref{sec:sd-main} and~\ref{sec:ward} work out the general theory of Schwinger-Dyson equations and Ward identities in NN-FT. Sections~\ref{sec:ml_examples} and~\ref{sec:examples} work through several illustrative examples in machine learning and physics. Specifically: 
\begin{itemize}
\item In \S\ref{sec:sd-main}, we derive the Schwinger--Dyson equations and the contracted identity from total derivatives in parameter space. The breaking function makes its first appearance.
\item In \S\ref{sec:ward}, we specialize the contracted identity to absorbed symmetries, introduce the relationship between the breaking function and symmetries, and relate parameter space currents to field space Ward identities.
\item In \S\ref{sec:mlp-symmetries}, we derive Ward identities for input translations, input rotations, and output space transformations in multilayer perceptrons, and identify the conditions on the parameter density under which each symmetry is preserved.
\item In \S\ref{sec:attention}, we apply the breaking function to the attention mechanism in transformers.
\item In \S\ref{sec:com_scalar}, we analyze a complex scalar with a $U(1)$ symmetry and show how a mass splitting produces explicit breaking whose parameter space insertion matches the field space breaking inside correlators.
\item In \S\ref{sec:scale-anomaly}, we study scale transformations.
The raw finite dimensional Jacobian in a cutoff architecture is not, by itself, the physical trace anomaly.
This is obtained after including boundary fluxes and renormalized density deformations, the parameter space Ward identity reproduces the one-loop $\phi^4$ beta function.
\item In \S\ref{sec:weyl}, we study the NN-FT for the bosonic string~\cite{Frank:2026bui} on an $S^2$ worldsheet.
We compute the Weyl anomaly from mode counting in parameter space and recover the critical dimension $D=26$.
\item In \S\ref{sec:higher-form}, we examine compact bosons and related topological sectors~\cite{Ferko:2026ken}.
Here, we explain how higher-form and discrete symmetries are represented by sector identities rather than by smooth vector fields on a single parameter space sheet.
\item In \S\ref{sec:T}, T-duality of the bosonic string worldsheet sigma model~\cite{Ferko:2026ken} becomes a finite, discrete analogue of the breaking function Ward identity.
\end{itemize}
We conclude in Section~\ref{sec:conclusions} and provide outlook for future work.

\section{Schwinger--Dyson Equations}
\label{sec:sd-main}

The Schwinger--Dyson equations are identities among correlation functions obtained by demanding that the generating functional $Z[J]$ is invariant under infinitesimal redefinitions of the integration variable.
Usually they are computed in the case that $Z[J]$ is represented by the Feynman path integral.
In this section we derive them in NN-FT, where $Z[J]$ is represented in parameter space.
The associated Schwinger--Dyson equations can be thought of as being induced by parameter space fluctuations, which determine field fluctuations via the neural network architecture.

The generating functional over network parameters $\theta \in \mathbb{R}^{|\theta|}$ is
\begin{equation}
\label{eq:partition}
Z[J] = \int \dd\theta\; p(\theta)\;
\exp\!\Bigl(\int \dd^d x \sum_{\alpha} J_\alpha(x)\, \phi_\theta^\alpha(x)\Bigr) \,,
\end{equation}
where $p(\theta) > 0$ is the density, $\phi_\theta^\alpha(x)$ is the network output ($\alpha = 1, \ldots, k$), and $J_\alpha(x)$ is an external source.
The correlators are parameter space expectations:
\begin{align}
\label{eq:correlator}
\vev{\phi^{\alpha_1}(x_1) \cdots \phi^{\alpha_n}(x_n)}
= \frac{1}{Z[0]} \int \dd\theta\; p(\theta)\;
\phi_\theta^{\alpha_1}(x_1) \cdots \phi_\theta^{\alpha_n}(x_n) \,.
\end{align}
We define the \emph{score function} $s_a(\theta) \equiv \partial_a \log p(\theta)$ and the source exponential $\mathcal{O}_J \equiv e^{\int \dd^d x\, \sum_\alpha J_\alpha \phi_\theta^\alpha}$, where $\partial_a \equiv \partial/\partial\theta_a$.
While $\langle s_a \rangle = 0$, the probability weighted sum of the square of the score function yields the Fisher information.
Indices $a, b, c$ label parameter space ($1, \ldots, |\theta|$); $i, j$ label input space ($1, \ldots, d$); $\alpha, \beta$ label output space ($1, \ldots, k$).
All sums are written explicitly.

Fix $a \in \{1, \ldots, |\theta|\}$ and let $\mathcal{O}(\theta, J)$ be any smooth function. The integral of a total derivative in $\theta_a$ over the parameter domain $\mathcal{M}_a$ gives, after expanding via the product rule and integrating over the remaining parameters,
\begin{equation}
\label{eq:SD}
\boxed{\vev{\partial_a \mathcal{O}}
= -\,\vev{s_a(\theta)\, \mathcal{O}}
+\frac{1}{Z[0]}\int \dd\theta\; \partial_a\bigl[p(\theta)\,\mathcal{O}\bigr]}\;,
\end{equation}
one identity per parameter $\theta_a$, valid for any smooth $\mathcal{O}$. The last term is the boundary contribution from the total derivative and vanishes when $p\,\mathcal{O}$ decays sufficiently at the boundary of $\mathcal{M}_a$.

Now let $\xi^a(\theta)$ be an arbitrary smooth vector field on parameter space. Replacing $\mathcal{O} \to \xi^a \mathcal{O}$ in~\eqref{eq:SD}, applying the product rule on the left, and summing over $a$ yields the \emph{contracted Schwinger--Dyson identity}:
\begin{equation}
\label{eq:contracted}
\boxed{\biggl\langle \sum_a \xi^a\, \partial_a \mathcal{O} \biggr\rangle
= -\,\vev{B(\theta)\, \mathcal{O}} +\frac{1}{Z[0]}\oint_{\partial\mathcal{M}}\dd S\;(\xi\cdot \hat{n})\,p(\theta)\,\mathcal{O}}\;,
\end{equation}
where the \emph{breaking function} is
\begin{equation}
\label{eq:B-def}
B(\theta) := \sum_a \xi^a\, s_a + \sum_a \partial_a \xi^a \,.
\end{equation}
No symmetry is assumed; this holds for any $\xi^a(\theta)$. The boundary term collects the individual boundary contributions from~\eqref{eq:SD} via the divergence theorem.

Field variations in NN-FT are determined by such parameter space vector fields $\xi^a(\theta)$ as
\begin{equation}
\delta_\xi \phi^\alpha(x) \equiv \sum_a \xi^a\, \partial_a \phi^\alpha(x) \,.
\end{equation}
Setting $\mathcal{O} = \mathcal{O}_J$ in~\eqref{eq:contracted}, functional differentiation with respect to $J_{\alpha_1},\dots,J_{\alpha_n}$ and setting $J=0$ yields the $n$-point form of the Schwinger--Dyson equations:
\begin{align}
\label{eq:npt}
\sum_{k=1}^{n} \bigl\langle
\phi^{\alpha_1}(x_1) \cdots \delta_\xi\phi^{\alpha_k}(x_k) \cdots \phi^{\alpha_n}(x_n)
\bigr\rangle
&= -\bigl\langle B\,
\phi^{\alpha_1}(x_1) \cdots \phi^{\alpha_n}(x_n) \bigr\rangle \notag\\
&\quad +\frac{1}{Z[0]}\oint_{\partial\mathcal{M}}dS\;(\xi\cdot\hat{n})\,p(\theta)\prod_{i=1}^n \phi^{\alpha_i}(x_i)\;,
\end{align}
which is similar to textbook forms of the Schwinger--Dyson equations in ordinary quantum field theory.

The breaking function $B(\theta)$ is related to a conserved current on parameter space.
Substituting $s_a = (\partial_a p)/p$ and using the product rule:
\begin{equation}
\label{eq:pB}
p\, B = \sum_a \partial_a(\xi^a\, p).
\end{equation}
Since $p > 0$, $B = 0$ if and only if
\begin{equation}
\label{eq:continuity}
\sum_a \partial_a\bigl(\xi^a\, p\bigr) = 0.
\end{equation}
This is a continuity equation: the probability current 
\begin{equation} 
    j^a = \xi^a\, p
\end{equation}
is divergence free.
We emphasize that the current lives in the $|\theta|$-dimensional parameter space and is \emph{not} a local current in spacetime.
It would be interesting to study its relevance for machine learning.

We observe that the logic behind our derivations is the same ``change of variables'' principle familiar from standard quantum field theory: because the field $\phi(x)$ is a dummy variable in the path integral, we can introduce a shift without affecting the result.
Exploring the consequences of this mathematical triviality is exactly how we derive the quantum equations of motion (Schwinger--Dyson equations) and quantum symmetry conservation laws (Ward identities).
Here, we have simply performed an identical operation for the parameters that define the NN-FT.

\section{Ward identities}
\label{sec:ward}

Ward identities arise from the Schwinger--Dyson equation by considering field variations that are symmetries and therefore leave $Z[J]$ invariant.
We will derive Ward identities in the parameter space description and see the relevance of the breaking function $B(\theta)$.

An infinitesimal field variation $\delta\phi_\theta$ induces a fluctuation in $Z[J]$:
\begin{equation}
\delta Z[J]
= \int \dd\theta\; p\,
\biggl(\int \dd^d x \sum_\alpha J_\alpha\, \delta\phi^\alpha_\theta\biggr)
\, e^{\int J \phi_\theta} \,.
\end{equation}
If the fluctuation is induced by movement in parameter space we have 
$\delta\phi^\alpha = \sum_a \xi^a \partial_a \phi^\alpha$ and the right hand side becomes $\langle \sum_a \xi^a\partial_a\mathcal{O}_J\rangle \cdot Z[0]$. From the contracted Schwinger--Dyson identity~\eqref{eq:contracted}, 
\begin{equation}
\label{eq:deltaZ}
\delta Z[J] = -\bigl\langle B(\theta)\, \mathcal{O}_J \bigr\rangle \cdot Z[0] +\oint_{\partial\mathcal{M}}dS\;(\xi\cdot n)\,p(\theta)\mathcal{O}_J\,,
\end{equation}
Of course, if the field variation is a symmetry we must have $\delta Z = 0$.
Equation~\eqref{eq:deltaZ} shows that in the absence of the boundary term, $B(\theta)=0$ is a sufficient condition for invariance of the partition function (in Section~\ref{sec:attention} we will see an example where $B\neq0$ but the symmetry still holds). In general however, there must be cancellation between the breaking function term and the boundary flux in order for the symmetry to hold. When $\delta Z=0$, we can take functional derivatives with respect to $J_{\alpha_1},\dots, J_{\alpha_n}$ and set $J=0$ to obtain
\begin{equation}
\label{eq:ward}
    \langle B\,\phi^{\alpha_1}(x_1)\cdots \phi^{\alpha_n}(x_n)\rangle=\frac{1}{Z[0]}\oint_{\partial\mathcal{M}}dS\;(\xi\cdot n)\,p(\theta)\prod_{i=1}^n \phi^{\alpha_i}(x_i)\;,
\end{equation}
the \emph{$n$-point parameter space Ward identity}.

The symmetry action by a group element $g\in G$ applies to all the network parameters as well as the input and the output, so we may write $\theta'$ as $\theta'(\theta,g)$; usually, $g$ acts on a subset of parameters.
In practice, global symmetries in NN-FT are often realized by the \emph{absorption mechanism}~\cite{Maiti:2021fpy}, whereby a field is transformed in a prescribed way and the transformation can be absorbed into a redefinition of parameters via the architecture equation.
These are akin to active (architecture) and passive (parameter density) transformations.

As an example, if in $\phi_\theta$ the first layer is a dense layer sending  $x_i \mapsto W_{ij} x_j + b_i$, then the rotation $R\in SO(d)$ may be absorbed as 
\begin{equation}
\phi_\theta(R \cdot x) =: \phi_{\theta'}(x) \,, \quad \text{where} \quad \theta' = (W, b) \mapsto (W R, b) \,.
\end{equation}
For a small rotation $R = I + \epsilon \omega$ we have $W_{ij}\to W_{ij}+\epsilon\,\delta W_{ij}$, with
\begin{equation}\label{eq:rotation_abs}
\delta W_{ij} = \sum_k W_{ik} \omega_{kj} =: \xi_{ij} \,, \quad \delta b_i = 0 \,.
\end{equation}
One could flatten the parameters into a vector $\theta_a$, but it is transparent to use a tensor $\xi_{ij}$.

\subsection{Score and Jacobian symmetry breaking}
\label{sec:score_jac}

Symmetry breaking is classified by the breaking function $B(\theta)$, which  decomposes as
\begin{equation}
\label{eq:B-decomp}
B(\theta) = \underbrace{\sum_a \xi^a\, s_a}_{\text{score}} + \underbrace{\sum_a \partial_a \xi^a}_{\text{Jacobian}}.
\end{equation}
The score term depends on the density through the score $s_a = \partial_a \log p$ and vanishes if the density is invariant along the flow generated by $\xi^a$.
The anomalous term depends only on the architecture determined generator $\xi^a$ and is independent of $p$. It is a finite-dimensional NN-FT analogue of the Fujikawa anomaly~\cite{Fujikawa1979}: the flat measure $\dd\theta$ is not invariant under $\theta \to \theta + \epsilon\xi(\theta)$ when $\sum_a \partial_a \xi^a \neq 0$, which arises from the Jacobian.

If both the score and the Jacobian terms vanish, $B=0$ and the continuity equation~\eqref{eq:continuity} holds.
It is a first order PDE for $p(\theta)$ given the architecture determined $\xi^a(\theta)$.
Its solutions characterize all densities that exactly preserve the symmetry.
Different architectures absorbing the same symmetry have different $\xi^a$ and therefore different constraints on $p$. 
In some examples, we will see that a deformation of an NN-FT may be absorbed equivalently into either a deformation of $p$ or of the architecture; where there is a symmetry, these choices leave $B=0$, but the score and Jacobian pieces may individually change.

Why have we not associated the score and Jacobian pieces to classical symmetry breaking and anomalies, respectively?
Indeed, it is a natural identification because in the path integral the analog of the score is 
\begin{equation}
    \frac{\delta}{\delta \phi} \log P[\phi] = - \frac{\delta S}{\delta \phi} \,,
\end{equation}
which is related to classical symmetries, and the analog of the Jacobian is a transformation of the measure $\mathcal{D}\phi$ associated to an anomaly.
However, a clean separation between the measure $\mathcal{D}\phi$ and the action is related to the existence of an $\hbar\to 0$ limit that separates classical effects from quantum effects.
Such a limit is generally absent in NN-FT, and furthermore different realizations of the same NN-FT may exhibit different decompositions of the breaking function into score and Jacobian terms.\footnote{For instance, using copula methods, the joint distribution of any set of random variables $\theta$ can be engineered by applying deterministic functions to a set of \emph{uniform} random variables $u$. By replacing each $\theta$ in an architecture $\phi_\theta$ with an appropriate $f(u)$, one can therefore formally realize the model using uniform parameters for which $s_a = 0$, but for which the breaking function $B$ should be unchanged. This illustrates the possibility of shifting contributions between the score and Jacobian terms by changing one's presentation of a NN-FT.}

\subsection{Local currents}
\label{sec:local-currents}

We have developed a formalism for Schwinger--Dyson equations and Ward identities in NN-FT, including a sharp understanding of the relationship of symmetries to the breaking function. 
In our treatment we have made no mention of a local current, as NN-FTs are in general non-local and the parameter space current $j_a$ knows nothing of spacetime. However, since they are universal for scalars, NN-FTs realizing theories with local symmetries should be able to represent their currents.

Specifically, Lagrangian field theories with continuous symmetries have a conserved Noether current
\begin{equation}
j^\mu(x) = \sum_\alpha
\frac{\partial \mathcal{L}}{\partial(\partial_\mu \phi^\alpha)}\,
\delta\phi^\alpha(x)\,,
\end{equation}
a vector field on the $d$-dimensional input space.
In NN-FT, another conserved quantity is the parameter space current $j^a = \xi^a\, p$ on $\mathbb{R}^{|\theta|}$.
The connection between the two proceeds through the action on field space.
The network map $\theta \mapsto \phi_\theta(\cdot)$ pushes the parameter space measure forward to a distribution on field configurations,
\begin{equation}
\label{eq:pfield}
p_{\text{field}}[\phi]
= \int \dd\theta\; p(\theta)\;
\delta\bigl[\phi - \phi_\theta\bigr]\,,
\end{equation}
defining a function space effective action
$S_{\text{eff}}[\phi] = -\log p_{\text{field}}[\phi]$;
we call this effective since it may be utilized when it is computable, even though it is not required in NN-FT.
Any observable depending on $\theta$ only through $\phi_\theta$ has identical expectation values under the parameter space and field space measures:
\begin{equation}
\label{eq:expect-match}
\vev{\mathcal{O}}_\theta
:= \frac{1}{Z[0]} \int \dd\theta\; p(\theta)\;
\mathcal{O}[\phi_\theta]
= \frac{1}{Z[0]}\int \mathcal{D}\phi\; p_{\text{field}}[\phi]\;
\mathcal{O}[\phi]
=: \vev{\mathcal{O}}_{\text{field}}\,,
\end{equation}
as follows by integrating the delta function in $p_{\text{field}}$ against $\mathcal{O}$.
In particular, all NN-FT correlators are identically the correlators of the effective field theory with action $S_{\text{eff}}$, which is difficult to compute in general.

The effective field theory has its own Schwinger--Dyson equation, obtained from the vanishing of a total functional derivative,
\begin{equation}
\int \mathcal{D}\phi\;
\frac{\delta}{\delta\phi^\alpha(x)}
\bigl[p_{\text{field}}[\phi]\,\mathcal{O}[\phi]\bigr] = 0\,.
\end{equation}
Contracting with the symmetry variation $\delta\phi^\alpha(x)$ by the same steps that led to the parameter space identity~\eqref{eq:contracted} yields a field space contracted Schwinger--Dyson identity:
\begin{equation}
\label{eq:field-ward}
\biggl\langle \sum_\alpha \int \dd^d x\;
\delta\phi^\alpha(x)\,
\frac{\delta \mathcal{O}}{\delta\phi^\alpha(x)}
\biggr\rangle_{\!\text{field}}
= -\bigl\langle B_{\text{eff}}[\phi]\,
\mathcal{O}[\phi] \bigr\rangle_{\text{field}}\,,
\end{equation}
where the field space breaking function is
\begin{equation}
\label{eq:Beff}
B_{\text{eff}}[\phi]
= -\sum_\alpha \int \dd^d x\;
\frac{\delta S_{\text{eff}}}{\delta \phi^\alpha(x)}\,
\delta\phi^\alpha(x)
+ \mathcal{A}_{\text{field}}\,,
\end{equation}
with
\begin{equation}
\mathcal{A}_{\text{field}}
= \sum_\alpha \int \dd^d x\;
\frac{\delta(\delta\phi^\alpha(x))}{\delta\phi^\alpha(x)}\,,
\end{equation}
the functional divergence of the field variation, equivalently the Fujikawa anomaly arising from the Jacobian of $\mathcal{D}\phi$ under $\phi \to \phi + \delta\phi$.

For an absorbed symmetry,
$\delta_\xi\phi^\alpha = \sum_a \xi^a\,\partial_a\phi^\alpha = \delta\phi^\alpha$. 
For any $\mathcal{O}$ depending on $\theta$ only through $\phi_\theta$, the functional chain rule then gives
$\sum_a \xi^a\,\partial_a \mathcal{O}[\phi_\theta] = \sum_\alpha \int \dd^d x\;\delta\phi^\alpha \frac{\delta\mathcal{O}}{\delta\phi^\alpha}$,
so the left hand sides of~\eqref{eq:contracted} and~\eqref{eq:field-ward} agree by~\eqref{eq:expect-match}.
Equating the right hand sides:
\begin{equation}
\label{eq:matching}
\boxed{
\bigl\langle B(\theta)\, \mathcal{O}[\phi_\theta]
\bigr\rangle_\theta
= \bigl\langle B_{\text{eff}}[\phi]\,
\mathcal{O}[\phi] \bigr\rangle_{\text{field}} \,.}
\end{equation}
Inserting $1 = \int\mathcal{D}\phi\;\delta[\phi - \phi_\theta]$ on the left hand side,
\begin{equation}
    \int d\theta\;p(\theta) B(\theta)\int \mathcal{D}\phi\; \delta[\phi-\phi_\theta]\,\mathcal{O}[\phi]
    =\int \mathcal{D}\phi\;p_{\text{field}}[\phi]\,B_{\text{eff}}[\phi]\mathcal{O}[\phi]\,.
\end{equation}
This yields 
\begin{equation}
\label{eq:Beff-avg}
B_{\text{eff}}[\phi] = \frac{\int \dd\theta\; p(\theta)\, B(\theta)\; \delta[\phi - \phi_\theta]}
{\int \dd\theta\; p(\theta)\; \delta[\phi - \phi_\theta]}\,.
\end{equation}
by swapping the order of integration (justified whenever $\langle |B\mathcal{O}|\rangle_\theta<\infty$) and stripping off $\mathcal{O}[\phi]$.

The interpretation of $B_\text{eff}[\phi]$ deserves comment:
it is the density weighted average of the parameter space breaking function over the fiber
$\{\theta : \phi_\theta = \phi\}$,
\textit{i.e.}, over all points in parameter space $\theta$ where $\phi_\theta$ is equal to a fixed field configuration $\phi$.
Put differently, the architecture $\phi_\theta$ is in general a many-to-one map from parameter space to function space, which is explicitly encoded in $B_\text{eff}[\phi]$.
Since many parameter configurations generically produce the same field configuration, the effective theory cannot resolve individual $B(\theta)$ values and sees only this fiber average.
Equation~\eqref{eq:Beff-avg} is exact at any width and for any architecture, but in the case that the architecture is one-to-one, so that there is a single $\theta$ such that $\phi_\theta=\phi$ for fixed $\phi$, then both the numerator and denominator of~\eqref{eq:Beff-avg} localize to the unique preimage and any Jacobian factors cancel in the ratio, giving $B_\text{eff}[\phi] = B(\theta)$.
This applies, \textit{e.g.}, when the architecture is the Borel isomorphism that guarantees universality~\cite{FerkoHalversonMutchler2026}.

Let us finally relate this to local theories. When $S_{\text{eff}}$ is local, \textit{i.e.}, expressible as $\int \dd^d x\, \mathcal{L}_{\text{eff}}$, and the field variation $\delta\phi^\alpha$ is a symmetry of the theory so that $\delta\mathcal{L}_{\text{eff}}=\partial_\mu K^\mu$, then the explicit piece of $B_{\text{eff}}$ is
\begin{equation}
-\sum_\alpha \int \dd^d x\; \frac{\delta S_{\text{eff}}}{\delta\phi^\alpha(x)}\, \delta\phi^\alpha(x) = \int \dd^d x\;\partial_\mu j^\mu_{\text{eff}}(x)\,,
\end{equation}
with the effective Noether current
\begin{equation}
\label{eq:jeff}
j^\mu_{\text{eff}}(x) = \sum_\alpha
\frac{\partial \mathcal{L}_{\text{eff}}}
{\partial(\partial_\mu \phi^\alpha)}\,
\delta\phi^\alpha(x)-K^\mu\,.
\end{equation}
This provides the connection between the parameter space current $j^a = \xi^a\, p$ on $\mathbb{R}^{|\theta|}$ and the spacetime current $j^\mu_{\text{eff}}(x)$ on $\mathbb{R}^d$.
The two are related by a three step chain:
\begin{equation} \label{eq:chain}
j^a
\;\xrightarrow{\;\text{div}\;}
p\, B(\theta)
\;\xrightarrow{\;\text{fiber avg}\;}
B_{\text{eff}}[\phi]
\;\xrightarrow{\;\text{locality}\;}
\partial_\mu j^\mu_{\text{eff}}(x)\,.
\end{equation}
The parameter space divergence $\sum_a \partial_a j^a$ yields $p\, B$ exactly;
the fiber average~\eqref{eq:Beff-avg} yields $B_{\text{eff}}$ exactly;
and locality of $S_{\text{eff}}$ decomposes $B_{\text{eff}}$ into the spacetime divergence
$\int \dd^d x\;\partial_\mu j^\mu_{\text{eff}}$ plus the Fujikawa anomaly $\mathcal{A}_{\text{field}}$.
The first two steps hold at any width and for any architecture; the third requires locality of $S_{\text{eff}}$, which is architecture dependent.

In conclusion: we see that NN-FT is robust enough to recover local currents, as expected from universality, via passing from the parameter space current to the local current through the breaking functions with appropriate fiber averaging.
In future work, it would be interesting to study $B_\text{eff}[\phi]$ itself from a function space perspective in non-local theories.

\section{Machine learning examples}
\label{sec:ml_examples}

Before proceeding to examples of physically relevant NN-FTs, we wish to apply our formalism to architectures commonly utilized in machine learning. We will study the multilayer perceptron (a.k.a.\ deep feedforward network) and the attention mechanism that is foundational in LLMs.

\subsection{Multilayer perceptron}
\label{sec:mlp-symmetries}
We now apply the formalism to multilayer perceptrons (MLPs).
An $L$-layer MLP with layer widths $n_0 = d$ and $n_1, \ldots, n_L = k$
computes
\begin{equation}
\label{eq:mlp}
\phi_\theta(x) = W^L \sigma\!\bigl(W^{L-1} \sigma\bigl(
\cdots \sigma(W^1 x + b^1) \cdots\bigr) + b^{L-1}\bigr) + b^L\,,
\end{equation}
where $W^\ell \in \mathbb{R}^{n_\ell \times n_{\ell-1}}$ and
$b^\ell \in \mathbb{R}^{n_\ell}$ are the weights and biases of
layer $\ell$, and $\sigma$ is the activation function applied
elementwise.
The parameters are
$\theta = \{W^\ell, b^\ell\}_{\ell=1}^L$,
with $|\theta| = \sum_\ell n_\ell(n_{\ell-1} + 1)$.

\paragraph{Spatial translations:}
The translation $x \mapsto x + \epsilon v$ enters the network only
through the first layer preactivation
$W^1 x + b^1 \mapsto W^1 x + (b^1 + \epsilon W^1 v)$,
so the generator is
\begin{equation}
\xi^{b^1_m} = \sum_j W^1_{mj}\, v^j\,,
\qquad \text{all other } \xi^a = 0\,.
\end{equation}
Since $\xi$ is constant in $b^1$, the Jacobian piece vanishes:
$\sum_a \partial_a \xi^a = 0$.
The breaking function is purely explicit,
\begin{equation}
\label{eq:B-trans}
B = \sum_{m,j} W^1_{mj}\, v_j\, s_{b^1_m}\,.
\end{equation}
For a Gaussian density $p \sim \mathcal{N}(0, \sigma^2 I)$,
$s_{b^1_m} = -b^1_m / \sigma^2$ and $B \neq 0$ generically:
the standard independent and identically distributed (i.i.d.)\ Gaussian density breaks translation invariance.

Stripping off the arbitrary direction $v^i$ gives $d$ independent
SD identities,
\begin{align}
\label{eq:ward-trans}
\sum_{k=1}^{n} \bigl\langle
\phi^{\alpha_1}(x_1) \cdots \partial_i\phi^{\alpha_k}(x_k)
\cdots \phi^{\alpha_n}(x_n) \bigr\rangle
= -\bigl\langle B_i\, \phi^{\alpha_1}(x_1)
\cdots \phi^{\alpha_n}(x_n) \bigr\rangle,
\end{align}
with $B_i = \sum_m W^1_{mi}\, s_{b^1_m}$.
When $B = 0$, the symmetry holds, and the two-point function satisfies
$(\partial_{x^i} + \partial_{y^i})\, G^{\alpha\beta}(x, y) = 0$,
forcing $G^{\alpha\beta}(x,y) = G^{\alpha\beta}(x - y)$.

The continuity equation~\eqref{eq:continuity} requires $p$ to be constant along the orbits $b^1 \mapsto b^1 + W^1 v$ for all $v \in \mathbb{R}^d$. For a factorized density $p(W^1)\,p(b^1)$ with generic $W^1$, this forces $p(b^1)$ to be constant in every direction, which is not normalizable on $\mathbb{R}^{n_1}$. This obstruction is avoided by periodic architectures. For example, in the random Fourier feature network~\cite{RahimiRecht2007,Halverson:2021aot}, the activation is $\cos(W\cdot x + c)$ with $c\sim\text{Unif}([-\pi,\pi])$, and the uniform density on the compact domain $[-\pi,\pi]$ is invariant under $c\mapsto (c + W\cdot v)\;\text{mod}\;2\pi$. This is the mechanism by which the scalar NN-FTs of \S\ref{sec:com_scalar} and \S\ref{sec:scale-anomaly} achieve translational invariance.

\paragraph{Spatial rotations:}
The rotation $x \mapsto (I + \epsilon\,\omega)\, x$ with antisymmetric
$\omega$ was given by~\eqref{eq:rotation_abs},
\begin{equation}
\xi^{W^1_{mj}} = \sum_k W^1_{mk}\, \omega_{kj}\,,
\qquad \text{all other } \xi^a = 0\,.
\end{equation}
The Jacobian piece is
$\sum_a \partial_a \xi^a = n_1\, \tr(\omega) = 0$
by antisymmetry.
The breaking function is
\begin{equation}
\label{eq:B-rot}
B = \sum_{m,j,k} W^1_{mk}\, \omega_{kj}\, s_{W^1_{mj}}\,.
\end{equation}
For a Gaussian i.i.d.\ density, $s_{W^1_{mj}} = -W^1_{mj}/\sigma^2$ and
\begin{equation}
B = -\frac{1}{\sigma^2} \sum_m
\sum_{j,k} W^1_{mj}\, W^1_{mk}\, \omega_{kj} = 0\,,
\end{equation}
since contracting the symmetric matrix
$\sum_m W^1_{mj} W^1_{mk}$ against the antisymmetric $\omega_{kj}$
vanishes identically.
The Gaussian i.i.d.\ density is rotation invariant.

There are $d(d{-}1)/2$ independent Schwinger--Dyson identities, one per
antisymmetric pair $i < j$, involving the angular momentum operator
$L_{ij} = x_i\partial_j - x_j\partial_i$:
\begin{align}
\label{eq:sd-rot}
\sum_{k=1}^{n} \bigl\langle
\phi^{\alpha_1}(x_1) \cdots L_{ij}\phi^{\alpha_k}(x_k)
\cdots \phi^{\alpha_n}(x_n) \bigr\rangle
= -\bigl\langle B_{ij}\, \phi^{\alpha_1}(x_1)
\cdots \phi^{\alpha_n}(x_n) \bigr\rangle\;.
\end{align}
More generally, a natural sufficient condition for rotation invariance is that $p(W^1)$ depends on $W^1$ only through the Gram matrix $G_{mn} = \sum_j W^1_{mj} W^1_{nj}$ (such as the i.i.d.\ Gaussian).

\paragraph{Internal (output space) Rotations:}
The output space transformation
$\phi_\alpha \to \sum_\beta (I + \epsilon\Lambda)_{\alpha\beta} \,\phi_\beta$
acts on the last layer parameters, which gives generators
\begin{equation}
\xi^{W^L_{\alpha m}} = \sum_\beta
\Lambda_{\alpha\beta}\, W^L_{\beta m}\,,
\quad
\xi^{b^L_\alpha} = \sum_\beta
\Lambda_{\alpha\beta}\, b^L_\beta\;,
\end{equation}
with all other $\xi^a=0$. The Jacobian piece is now nonzero in general:
\begin{equation}
\label{eq:anom-output}
\sum_a \partial_a \xi^a = (n_{L-1} + 1)\, \tr(\Lambda)\,.
\end{equation}
This vanishes for traceless generators, \textit{e.g.}, the Lie algebras
$\mathfrak{so}(k)$ and $\mathfrak{sl}(k,\mathbb{R})$, but not for scaling
transformations $\Lambda = \lambda\, I$, where
the Jacobian piece is $(n_{L-1} + 1)\, k\, \lambda$.

The full breaking function is
\begin{equation}
\label{eq:B-output}
B = \sum_{\alpha,\beta}\Bigl(
\sum_m \Lambda_{\alpha\beta}\, W^L_{\beta m}\, s_{W^L_{\alpha m}}
+ \Lambda_{\alpha\beta}\, b^L_\beta\, s_{b^L_\alpha}\Bigr)
+ (n_{L-1} + 1)\, \tr(\Lambda)\,.
\end{equation}
For SO($k$) transformations, $\Lambda$ is antisymmetric, so $\tr\Lambda=0$, and the Jacobian piece vanishes. The explicit piece also vanishes for a Gaussian i.i.d.\ density by the same antisymmetric/symmetric argument as for rotations, giving $B=0$. The Schwinger--Dyson identity is
\begin{align}
\label{eq:ward-output}
\sum_{k=1}^{n} \bigl\langle
\phi_{\alpha_1}(x_1) \cdots
\Lambda_{\alpha_k \beta}\, \phi_\beta(x_k)
\cdots \phi_{\alpha_n}(x_n) \bigr\rangle
= -\bigl\langle B\, \phi_{\alpha_1}(x_1)
\cdots \phi_{\alpha_n}(x_n) \bigr\rangle.
\end{align}
As a concrete consequence, when $B=0$ and SO($k$) is unbroken, the $n=2$ Ward identity along with Schur's lemma ensure that $G_{\alpha\beta}(x,y)=\delta_{\alpha\beta}\,g(x,y)$: all output channels share the same two-point function, and distinct channels are uncorrelated. This result holds for any architecture that absorbs SO($k$) with an invariant density. A natural sufficient condition for SO($k$) invariance is that
the last layer density depends on $(W^L, b^L)$ only through the
combinations $(W^L)^T W^L$, $(W^L)^T b^L$, and $\|b^L\|^2$.
The Gaussian i.i.d.\ density satisfies this condition.

\subsection{Attention mechanism}
\label{sec:attention}

We mentioned around equation (\ref{eq:deltaZ}) that, in the absence of boundary terms, $B = 0$ is a \emph{sufficient} condition to ensure $\delta Z = 0$. However, it is not \emph{necessary}: again ignoring the boundary term, the right side of (\ref{eq:deltaZ}) vanishes so long as $\langle B ( \theta) \mathcal{O}_J \rangle = 0$. One might ask whether this is possible; in this section we will show that the answer is affirmative, and give an example where $\langle B ( \theta ) \mathcal{O} \rangle = 0$ for any $\mathcal{O}$. The reason is that the symmetry we will consider here is viewed from the NN-FT perspective as a parameter redundancy, much like a gauge ``symmetry.'' The purpose of this section is to show that the breaking function machinery correctly recognizes this transformation as a redundant one and yields a trivial Ward identity, just as the Noether current associated with a local gauge symmetry is trivial.\footnote{More precisely, although Noether's theorem does produce a conserved current $j^\mu$ associated with $A_\mu \to A_\mu + \partial_\mu \lambda ( x )$, the resulting conserved charge acts trivially on all gauge-invariant observables.} This does not mean the symmetry is irrelevant; it means that its content lives in the geometry of parameter space rather than in the transformation properties of the realized field, and thus does not constrain correlators.

To emphasize the distinction, recall that the symmetries studied in \S\ref{sec:mlp-symmetries} are absorbed symmetries: a transformation of input or output space is compensated by a reparameterization $\theta\to\theta'$, with both field and parameters moving. These represent genuine symmetries of the NN-FT, a non-trivial map on the parameters that leaves the network \emph{ensemble} invariant.

The setting for this gauge redundancy analysis is the attention mechanism of transformers\footnote{Our goal here differs from that of~\cite{Ageev:2026qyh}, which instead realized the free scalar field using a transformer.}~\cite{Vaswani2017}, which contains a qualitatively different kind of symmetry in which the output is \emph{exactly} invariant under a family of parameter transformations, so that the parameters move but the field does not. This was identified in~\cite{Silverstein2026,Zhang2025} as a continuous rotational symmetry that shapes optimization dynamics.

Recall that, in the attention mechanism, a single self-attention head maps a sequence of input token vectors $x_1, \ldots, x_T \in \mathbb{R}^d$ to output vectors $\phi_1, \ldots, \phi_T \in \mathbb{R}^d$ via
\begin{equation}
\label{eq:attn-def}
\phi_i = W_O \sum_j \frac{\exp(x_i^T W_Q^T W_K x_j)}{\sum_m \exp(x_i^T W_Q^T W_K x_m)}\, W_V x_j \,,
\end{equation}
where $W_Q, W_K \in \mathbb{R}^{d_k \times d}$, $W_V \in \mathbb{R}^{d_v \times d}$, and $W_O \in \mathbb{R}^{d \times d_v}$.
The output $\phi_i$ depends on $W_Q$ and $W_K$ only through the product $M := W_Q^T W_K \in \mathbb{R}^{d \times d}$.
Under the joint rotation~\cite{Silverstein2026,Zhang2025}
\begin{equation}
\label{eq:QK-rotation}
W_Q \longrightarrow R\, W_Q\,,\qquad W_K \longrightarrow R\, W_K\,,\qquad R \in O(d_k)\,,
\end{equation}
with all other parameters unchanged, the product $M \to W_Q^T R^T R\, W_K = W_Q^T W_K = M$ is invariant, and therefore $\phi_i$ is unchanged.
An analogous $O(d_v)$ symmetry acts in the value--output sector~\cite{Silverstein2026}: $W_V \mapsto \widetilde{R}\, W_V$, $W_O \mapsto W_O\, \widetilde{R}^T$ with $\widetilde{R} \in O(d_v)$.
We focus on the QK sector below; the OV analysis is identical.

For the infinitesimal rotation $R = I + \epsilon\omega$, the generator is
\begin{equation}
\label{eq:xi-QK}
\xi^{(W_Q)_{ij}} = \sum_k \omega_{ik}\, (W_Q)_{kj}\,,
\qquad
\xi^{(W_K)_{ij}} = \sum_k \omega_{ik}\, (W_K)_{kj}\,,
\end{equation}
with all other $\xi^a = 0$.
We now show that $\delta_\xi \phi_i = 0$ explicitly.
Since $\phi_i$ depends on $W_Q$ and $W_K$ only through $M_{ab} = \sum_k (W_Q)_{ka} (W_K)_{kb}$, the chain rule gives
\begin{equation}
\frac{\partial \phi_i}{\partial (W_Q)_{pq}} = \sum_b \frac{\partial \phi_i}{\partial M_{qb}}\, (W_K)_{pb}\,, \qquad
\frac{\partial \phi_i}{\partial (W_K)_{pq}} = \sum_a \frac{\partial \phi_i}{\partial M_{aq}}\, (W_Q)_{pa}\,.
\end{equation}
The field variation splits into two terms, $\delta_\xi \phi_i = T_1 + T_2$, with
\begin{equation}
T_1 = \sum_{q,b} \frac{\partial \phi_i}{\partial M_{qb}} \sum_{p,k} \omega_{pk}\, (W_Q)_{kq}\, (W_K)_{pb}\,, \qquad
T_2 = \sum_{a,q} \frac{\partial \phi_i}{\partial M_{aq}} \sum_{p,k} \omega_{pk}\, (W_K)_{kq}\, (W_Q)_{pa}\,.
\end{equation}
In $T_1$, swapping the dummy indices $p \leftrightarrow k$ and using $\omega_{kp} = -\omega_{pk}$ gives $T_1 = -T_2$, so
\begin{equation}
\label{eq:exact-invariance}
\delta_\xi \phi_i = 0 \qquad \text{for all } \theta\,.
\end{equation}
This is the infinitesimal version of $R^T R = I$.
Unlike the absorbed symmetries of \S\ref{sec:mlp-symmetries}, there is no compensating transformation of the field: the output is simply invariant.
Again, the $O(d_k)$ rotation is a redundancy of the parameterization, which is why we refer to it as the analogue of a gauge symmetry from the perspective of NN-FT correlators.

We now test whether the breaking function formalism correctly identifies this gauge redundancy as such. The Jacobian piece of the breaking function vanishes by antisymmetry of $\omega$: $\sum_a \partial_a \xi^a = 2d\,\tr(\omega) = 0$.
The breaking function is therefore purely the score piece, $B = \sum_a \xi^a s_a$, which vanishes for any $O(d_k)$-invariant density such as the Gaussian i.i.d.\ by the same antisymmetric/symmetric contraction as in \S\ref{sec:mlp-symmetries}, but is generically nonzero for a non-invariant density. 

Nevertheless, since $\delta_\xi\phi=0$ even for non-invariant densities, correlators involving $\phi$ are also invariant and in particular
$\langle B \mathcal{O}[\phi_i]\rangle=0$ for any $\mathcal{O} [\phi_i]$.
To see this, decompose parameter space into directions along and transverse to the $O(d_k)$ orbits, $\theta = (\theta_\parallel, \theta_\perp)$.
Since $\delta_\xi\phi = 0$, any observable $\mathcal{O}[\phi_\theta]$ depends only on $\theta_\perp$, while $B = \xi^a s_a$ is a derivative of $\log p$ along $\theta_\parallel$.
The expectation value factorizes as
\begin{equation}
\label{eq:orbits}
\langle B\,\mathcal{O}\rangle = \int d\theta_\perp\, \mathcal{O}(\theta_\perp) \int_{\text{orbit}} d\theta_\parallel\; \underbrace{\xi^a \partial_a p}_{=\;\partial_a(\xi^a p)} = 0\,,
\end{equation}
where the inner integral vanishes by the divergence theorem on each compact boundaryless orbit of the compact Lie group $O(d_k)$.
The $n$-point Schwinger--Dyson equation~\eqref{eq:npt} therefore reduces to $0 = 0$: the left side vanishes because $\delta_\xi\phi_i = 0$, and the right side by~\eqref{eq:orbits}. By contrast, for an absorbed symmetry such as the MLP rotation, the left side involves the nontrivial field space operation $\sum_k \langle \phi \cdots L_{ij}\phi_k \cdots \phi \rangle$, and the vanishing of the right side forces the angular momentum Ward identities~\eqref{eq:sd-rot}. Although the breaking function is not a charge, this behavior is again reminiscent of the would-be charge associated with gauge transformations in ordinary field theory: for a local gauge transformation with infinitesimal gauge parameter $\lambda = \lambda ( x )$ and Noether charge $Q_\lambda$, one has $\langle Q_\lambda \mathcal{O} \rangle = 0$ for all gauge-invariant operators $\mathcal{O}$.

One can also interpret this calculation as the statement that the effective breaking function $B_{\text{eff}}$ vanishes. Since $\delta_\xi\phi_i = 0$, the functional delta function $\delta[\phi - \phi_\theta]$ in the numerator of~\eqref{eq:Beff-avg} is constant along the flow of $\xi$, so the integrand $p\,B\,\delta[\phi-\phi_\theta] = \sum_a\partial_a(\xi^a p\,\delta[\phi-\phi_\theta])$ is a total divergence and integrates to zero on the compact boundaryless $O(d_k)$ orbits. Thus, while the raw breaking function $B$ is generically non-zero, its density weighted fiber average vanishes: $B_{\text{eff}} = 0$. 

To conclude, attention gives a clean example where the breaking function formalism distinguishes a physical symmetry from a parameter redundancy. Although the rotational symmetry discussed here may have implications for optimization dynamics~\cite{Silverstein2026,Zhang2025}, from the perspective of NN-FT correlation functions, it is merely a gauge redundancy: $\delta_\xi \phi_i = 0$ and, while $B \neq 0$, one finds that $B_{\text{eff}} = 0$ and the Ward identity does not offer any new constraints. It would be interesting to engineer an NN-FT presentation of Maxwell theory and repeat the analysis here for gauge transformations to see whether the same structure emerges.

\section{Physics examples}
\label{sec:examples}

We will now study our formalism in the context of a number of physical NN-FTs: a $U(1)$ symmetry for the complex scalar, a scale anomaly in $\phi^4$ theory, the Weyl anomaly and critical dimension in the bosonic string, higher-form symmetries, and T-duality.

\subsection{$U(1)$ and complex scalar}\label{sec:com_scalar}

Consider two real scalar architectures~\cite{halverson2021building, demirtas2023neural} indexed by~$\ell$
\begin{equation}
    \phi_\ell(x)=\frac{C}{\sqrt{N}}\sum_{j=1}^N \frac{a_j^{(\ell)}}{\sqrt{b_j^2+m_\ell^2}}\cos(b_j\cdot x+c_j)\;,
    \label{eqn:real_scalar_arch}
\end{equation}
where $a_j^{(1)}$ and $a_j^{(2)}$ are independent real Gaussians of the same variance $\sigma^2$, $b_j$ is uniformly drawn from $B^d_\Lambda$, the $d$-ball of radius $\Lambda$, $c_j$ is uniformly drawn from the interval $[-\pi,\pi]$, and $C$ is a density dependent normalization constant. All distributions are i.i.d.\ along the $j$ index. Both $\phi_1$ and $\phi_2$ share the same frequencies $b_j$ and phases $c_j$. It is known that~\cite{halverson2021building}
\begin{equation}
    \langle\phi_\ell\phi_\ell\rangle=\frac{\chi_{|p|<\Lambda}}{p^2+m_\ell^2}\;,
\end{equation}
where $\chi$ denotes an indicator function that restricts momenta to $B^d_\Lambda$. Furthermore, $\langle \phi_1\phi_2\rangle=0$ by independence of the Gaussian output parameters.

Define hyperparameters $m$ and $\varepsilon$ such that
\begin{equation}
    m_1^2=m^2+2\varepsilon\;,\qquad m_2^2=m^2-2\varepsilon\;.
\end{equation}
Consider the case of $\varepsilon=0$, so $m_1^2=m_2^2=m^2$. The infinite width effective action is
\begin{align}
    S^{\text{eff}}_0&=\int d^dx\;\Big(\frac{1}{2}\partial^\mu \phi_1\partial_\mu\phi_1+\frac{1}{2}\partial^\mu\phi_2\partial_\mu\phi_2+\frac{1}{2}m^2(\phi_1^2+\phi_2^2)\Big)\notag\\
    &:=\int d^dx\;\Big(\frac{1}{2}\partial^\mu\phi\partial_\mu\phi^*+\frac{1}{2}m^2|\phi|^2\Big)\;,\label{eq:free_eff_action}
\end{align}
where we have defined the \textit{free complex scalar} $\phi:=\phi_1+i\phi_2$. This theory has a symmetry under the $U(1)$ transformation $\phi\mapsto e^{i\alpha}\phi$, which manifests in the free fields $\phi_{1,2}$ as $(\phi_1, \phi_2)\mapsto(\phi_1\cos\alpha-\phi_2\sin\alpha, \phi_1\sin\alpha+\phi_2\cos\alpha)$. The field space generators are
\begin{equation}
    \delta\phi_1=-\phi_2\;,\qquad\delta\phi_2=\phi_1\;,\label{eq:scalar_u1}
\end{equation}
so that the actual transformation is $\phi_\ell \mapsto \phi_\ell + \alpha\,\delta\phi_\ell$. It is straightforward to check that $B=0$ in this case.

Now consider $\varepsilon\neq0$ (so $m_1\neq m_2$). The effective action in terms of the complex scalar $\phi$ is
\begin{equation}
    S^{\text{eff}}_\varepsilon=\int d^dx\;\Big[\frac{1}{2}\partial^\mu\phi\partial_\mu\phi^*+\frac{1}{2}m^2|\phi|^2+\frac{1}{2}\varepsilon(\phi\phi+\phi^*\phi^*)\Big]\;.\label{eq:broken_sym_action}
\end{equation}
This effective action describes a theory of a complex scalar with mass $m$ and a deformation proportional to $\varepsilon$ that breaks the $U(1)$ symmetry of~\eqref{eq:free_eff_action}. Note, however, that there is not a simple architecture for the complex scalar of mass $m$ in this symmetry breaking case: it is easiest to describe the system using two free scalars of different masses.

Even though~\eqref{eq:broken_sym_action} breaks $U(1)$ symmetry, we will still absorb the $U(1)$ transformation as in~\eqref{eq:scalar_u1} in order to see an example of explicit symmetry breaking. Absorbing the generators~\eqref{eq:scalar_u1} into parameter space requires
\begin{equation}
    \frac{\xi^{a_j^{(1)}}}{\sqrt{b_j^2+m_1^2}}=-\frac{a_j^{(2)}}{\sqrt{b_j^2+m_2^2}}\,,
\qquad \frac{\xi^{a_j^{(2)}}}{\sqrt{b_j^2+m_2^2}}=\frac{a_j^{(1)}}{\sqrt{b_j^2+m_1^2}}\,.
\end{equation}
Define the $b_j$-dependent ratio $r_j$ by
\begin{equation}
    r_j:=\sqrt{\frac{b_j^2+m_1^2}{b_j^2+m_2^2}}
\end{equation}
such that the parameter space generators are
\begin{equation}
    \xi^{a_j^{(1)}}=-r_j \,a_j^{(2)}\;,\qquad \xi^{a_j^{(2)}}=r_j^{-1}\,a_j^{(1)}\;.
\end{equation}
It is immediately clear that there is no anomaly term in the breaking function $B$ that describes this system: $\partial_{a^{(1)}}\xi^{a^{(1)}}=\partial_{a^{(2)}}\xi^{a^{(2)}}=0$. However, we have
\begin{equation}
    B=B_{\text{explicit}}=\frac{1}{\sigma^2}\sum_j a_j^{(1)}a_j^{(2)}(r_j-r_j^{-1})
    =\frac{4\varepsilon}{\sigma^2}\sum_j \frac{a_j^{(1)} a_j^{(2)}}{\sqrt{(b_j^2+m_1^2)(b_j^2+m_2^2)}}\;,
\end{equation}
an explicit symmetry breaking term.

We wish to compare this parameter space breaking function to the field space breaking function.
Since $\mathcal{A}_{\text{field}} = 0$, we have
$B_{\text{eff}}[\phi] = -\delta S^{\text{eff}}_\varepsilon[\phi]$,
where $\delta S$ denotes the variation of the action under
the generators~\eqref{eq:scalar_u1}.
Checking the fiber averaged relation~\eqref{eq:Beff-avg}
directly is difficult, as the fiber
$\{\theta : \phi_\theta = \phi\}$, the set of all parameter
configurations producing a given field, is hard to
characterize explicitly for the random Fourier feature
architecture.

However, we can verify that the two breaking functions agree
inside correlators, as in~\eqref{eq:matching}.
For $\mathcal{O} = \phi_1(x)\phi_2(y)$,
the field space side is
\begin{equation}
\bigl\langle B_{\text{eff}}\,
\phi_1(x)\phi_2(y)\bigr\rangle_{\text{field}}
= 4\varepsilon \int d^d z\;
\bigl\langle \phi_1(z)\phi_2(z)\phi_1(x)\phi_2(y)
\bigr\rangle\,.
\end{equation}
Since $\langle\phi_1\phi_2\rangle = 0$, the only Wick contraction
is $\langle\phi_1(z)\phi_1(x)\rangle
\langle\phi_2(z)\phi_2(y)\rangle
= G_1(z{-}x)\, G_2(z{-}y)$.
In momentum space the $z$-integral collapses the two momenta,
giving
\begin{equation}
\label{eq:Beff-corr}
\bigl\langle B_{\text{eff}}\,
\phi_1(x)\phi_2(y)\bigr\rangle_{\text{field}}
= \frac{4\varepsilon}{(2\pi)^d}
\int_{B^d_\Lambda} d^d p\;
\frac{\cos\bigl(p\cdot(x{-}y)\bigr)}
{(p^2 + m_1^2)(p^2 + m_2^2)}\,.
\end{equation}
On the parameter side,
$\langle B\,\phi_1(x)\phi_2(y)\rangle_\theta$
involves a triple sum over neuron indices $(j,k,l)$
from $B$, $\phi_1$, and $\phi_2$ respectively.
Since $a^{(1)}$ and $a^{(2)}$ are independent,
the expectation factorizes into separate contractions
for each species.
The key contraction is
$\langle s_{a_j^{(\ell)}}\, a_k^{(\ell)} \rangle$:
for $j \neq k$, independence along $j$ gives
$\langle s_{a_j^{(\ell)}} \rangle
\langle a_k^{(\ell)} \rangle = 0$;
for $j = k$, the Schwinger--Dyson
equation~\eqref{eq:SD} with $\mathcal{O} = a$ gives
$\langle s_a \cdot a \rangle
= -\langle \partial_a a \rangle = -1$,
with no assumption on the distribution of $a$
beyond normalizability. Then, the remaining sum cancels with the $1/N$ factor by independence. Finally, by averaging over $c_j$ and writing the $b_j$ expectation as an integral, we obtain
\begin{equation}
\label{eq:B-corr}
\bigl\langle B\,\phi_1(x)\phi_2(y)
\bigr\rangle_\theta
= \frac{4\varepsilon}{(2\pi)^d}
\int_{B^d_\Lambda} d^d p\;
\frac{\cos\bigl(p\cdot(x{-}y)\bigr)}
{(p^2 + m_1^2)(p^2 + m_2^2)}\,,
\end{equation}
confirming~\eqref{eq:matching}. This result extends to any insertion $\mathcal{O}$ built from
products of $\phi_\ell$.

\paragraph{Parameter density deformation:}
\label{sec:dens_def}
Let $p_G=p(a^{(1)})p(a^{(2)})p(b)p(c)$ be the parameter density of the free complex scalar (equivalently of the two real scalars with the same mass $m$). Instead of deforming the masses, the action~\eqref{eq:broken_sym_action} can be obtained via deforming $p_G$ as
\begin{equation}
    p_G\mapsto p_\varepsilon=p_G\,e^{-V}\;,
\end{equation}
with
\begin{equation}
    V=\frac{\varepsilon}{2}\int d^dx\;[\phi\phi+\phi^*\phi^*]\;.
\end{equation}
Such a density deformation changes the scores, so the breaking function changes as a result. The Jacobian part of the breaking function is still zero since the architecture is unaffected, but the explicit breaking function changes. The generators are
\begin{equation}
    \xi^{a_j^{(1)}}=-a_j^{(2)}\;,\qquad \xi^{a_j^{(2)}}=a_j^{(1)}\;,
\end{equation}
and the change in the scores relative to the free case is
\begin{equation}
    \Delta s_a=-\partial_aV\;.
\end{equation}
Then, because $B=0$ in the free case, we have
\begin{align}
    B=\Delta B&=\sum_a \xi^a\Delta s_a\notag\\
    &=\sum_j\Big[-a_j^{(2)}\big(-\partial_{a_j^{(1)}}V\big)+a_j^{(1)}\big(-\partial_{a_j^{(2)}}V\big)\Big]\notag\\
    &=4\varepsilon\int d^dx\;\phi_1(x)\phi_2(x)=B_{\text{eff}}[\phi_1,\phi_2]\;,
\end{align}
demonstrating that two different mechanisms, architecture deformation and density deformation, can lead to two different expressions for the breaking functions. Despite this, the breaking functions are equal when acting on observables within expectation values, as in~\eqref{eq:matching}. 

\subsection{Scale anomaly}
\label{sec:scale-anomaly}

We will now demonstrate a well-known scale anomaly in NN-FT, in massless $\phi^4$ theory in $4d$, from one-loop violation of the Ward identity. In doing so, we recover the one-loop $\beta$-function.

\paragraph{Massless free scalar:}
Consider the massless real scalar architecture obtained from~\eqref{eqn:real_scalar_arch} by setting the mass to zero and restricting to one real field.
In the infinite width limit its canonically normalized effective action is
\begin{equation}
\label{eq:effect_action}
S_{0}^{\text{eff}} = \int \dd^4x\;\frac{1}{2}\partial_\mu\phi\,\partial^\mu\phi\,.
\end{equation}
This theory is classically scale invariant.
The corresponding field transformation is
\begin{equation}
\phi(x) \longrightarrow \alpha\,\phi(\alpha x) \,,
\qquad \Delta^{\text{cl}}_\phi=1 \,,
\end{equation}
and it is absorbed by the parameter space transformation
\begin{equation}
\label{eq:scale-parameter flow}
a_j\longrightarrow \alpha^2 a_j \,,
\qquad
b_{j\mu}\longrightarrow \alpha b_{j\mu} \,,
\qquad
c_j\longrightarrow c_j \,.
\end{equation}
Equivalently, for $\alpha=1+\epsilon$, the nonzero components of the vector field are
\begin{equation}
\xi^{a_j}=2a_j \,,
\qquad
\xi^{b_{j\mu}}=b_{j\mu} \,.
\end{equation}
For the Gaussian density of the output weights and the uniform density of the frequencies in the interior of $B^4_\Lambda$, the breaking function is
\begin{equation}
\label{eq:free-scale-breaking}
B  = \xi\cdot s+\partial\cdot\xi = -\frac{2}{\sigma^2}\sum_j a_j^2 + (d+2)N \,,
\quad d=4 \,,
\end{equation}
so that the Jacobian piece is $6N$. As emphasized in Section~\ref{sec:score_jac}, a non-zero Jacobian piece of the breaking function should not necessarily be considered an anomaly.

\paragraph{Free theory Ward identity:}
We now compute the free theory Ward identity for the sake of illustration. In the limit that the cutoff is taken to infinity, it should be satisfied, but should not for finite cutoff. We will see this explicitly. By comparison, in the interacting theory we will see a violation of the Ward identity that is not simply a cutoff artifact.

Since the frequency variables have compact support, we must retain the
boundary term in~\eqref{eq:ward}.  Only the components
$|b_j|=\Lambda$ contribute because $\xi^{c_j}=0$.  Taking polynomial combinations of the fields in~\eqref{eq:ward} the scale Ward
identity holds if and only if the \emph{Ward violating term}
\begin{equation}
\label{eq:B0-def}
\mathcal{B}^{(0)}[\mathcal{O}]
:= -\langle B\,\mathcal{O}\rangle
   + F_0[\mathcal{O}]
\end{equation}
vanishes for every polynomial observable $\mathcal{O}$, where
\begin{equation}
\label{eq:flux}
    F_0[\mathcal{O}]=\frac{1}{Z_0}\int d\theta\;\sum_{j,\mu}\partial_{b_{j\mu}}(b_{j\mu}\,p\,\mathcal{O})
\end{equation}
is the boundary flux term, and $Z_0$ is the free partition function (conventionally chosen to be $1$). We evaluate $\mathcal{B}^{(0)}$ on the four-point observable
$\mathcal{O}_4 := \prod_{r=1}^4\phi(x_r)$; the same observable
appears in the interacting theory, where the analogous
computation yields the beta function. 

The observable $\mathcal{O}_4$ is homogeneous of degree $4$ in the
Gaussian coefficients $a_j$, so Euler's theorem gives
$\sum_j a_j\,\partial_{a_j}\mathcal{O}_4 = 4\,\mathcal{O}_4$.
Using Gaussian integration by parts,
$\langle a_j^2\,f\rangle = \sigma^2\langle f\rangle
+ \sigma^2\langle a_j\,\partial_{a_j}f\rangle$,
and summing over $j$,
\begin{equation}
\sum_{j=1}^N
\langle a_j^2\,\mathcal{O}_4\rangle
= N\sigma^2\langle\mathcal{O}_4\rangle
+ \sigma^2\sum_{j=1}^N\langle a_j\,\partial_{a_j}\mathcal{O}_4\rangle
= \sigma^2(N+4)\langle\mathcal{O}_4\rangle\,,
\end{equation}
Hence, using the fact that $B = -\frac{2}{\sigma^2}\sum_j a_j^2 + 6N$,
\begin{equation}
-\langle B\,\mathcal{O}_4\rangle
= (8-4N)\,\langle\mathcal{O}_4\rangle\,.
\end{equation}

Expanding the derivative in the boundary flux~\eqref{eq:flux},
\begin{align}
F_0[\mathcal{O}_4]
&= \frac{1}{Z_0}\sum_{j,\mu}
   \int d\theta\;\partial_{b_{j\mu}}
   \bigl(b_{j\mu}\,p\,\mathcal{O}_4\bigr)
   \notag\\
&= \frac{1}{Z_0}\sum_{j,\mu}
   \int d\theta\;\Bigl[
   (\partial_{b_{j\mu}}b_{j\mu})\,
   p\,\mathcal{O}_4
   + b_{j\mu}\,(\partial_{b_{j\mu}}p)\,\mathcal{O}_4
   + b_{j\mu}\,p\,\partial_{b_{j\mu}}\mathcal{O}_4
   \Bigr]
   \notag\\
&= 4N\,\langle\mathcal{O}_4\rangle
   - \langle D_b\,\mathcal{O}_4\rangle\,,
\end{align}
where the score term $\partial_{b_{j\mu}}p$ vanishes in
the interior of $B^4_\Lambda$ because the frequency
density is uniform, and
$D_b := -\sum_{j,\mu} b_{j\mu}\,\partial_{b_{j\mu}}$
is the frequency space dilatation operator.  For the massless architecture,
$D_b\phi_\theta(x) = \phi_\theta(x) - x^\mu\partial_\mu\phi_\theta(x)$,
so by the product rule,
$D_b\mathcal{O}_4
= (4 - \sum_{r=1}^4 x_r^\mu\partial_{x_r^\mu})\mathcal{O}_4$.
Therefore
\begin{equation}
F_0[\mathcal{O}_4]
= (4N-4)\,\langle\mathcal{O}_4\rangle
  + \sum_{r=1}^4 x_r^\mu\partial_{x_r^\mu}
    \langle\mathcal{O}_4\rangle\,.
\end{equation}

Adding the bulk and flux pieces, the extensive $O(N)$ terms cancel:
\begin{equation}
\label{eq:B0-result}
\mathcal{B}^{(0)}[\mathcal{O}_4]
= \left(4 + \sum_{r=1}^4 x_r^\mu\partial_{x_r^\mu}\right)
  \langle\mathcal{O}_4\rangle\,.
\end{equation}
This is the position space dilatation operator acting on the free
four-point function.

In momentum space, the NN-FT propagator is
$G_0(p) = \theta(\Lambda-|p|)/p^2$, and the momentum space dilatation
acting on a single propagator gives
\begin{equation}
\label{eq:D2G0-mom}
-(2+p^\mu\partial_{p^\mu})\frac{\theta(\Lambda-|p|)}{p^2}
= \frac{\delta(|p|-\Lambda)}{\Lambda}\,.
\end{equation}
As $N\to\infty$, the free four-point function is purely disconnected,
\begin{equation}
    \langle\mathcal{O}_4\rangle
= G_0(p_1)G_0(p_2)\,\delta^4(p_1{+}p_2)\,\delta^4(p_3{+}p_4) + \text{2 perms}\;,
\end{equation}
so by the product rule every term in the Fourier transform
of~\eqref{eq:B0-result} contains at least one factor
of $\delta(|p_r|-\Lambda)$. Therefore
\begin{equation}
\label{eq:B0-vanish}
\mathcal{B}^{(0)}[\mathcal{O}_4]\Big|_{|p_r|<\Lambda} = 0\,,
\end{equation}
and the free theory Ward violation is supported entirely on the cutoff
shell and vanishes as $\Lambda\to\infty$ at fixed external momenta.

As we now
show, the interacting theory is qualitatively different: the one-loop
correction to the four-point function produces a logarithmic momentum dependence whose
dilatation is supported in the interior and survives
$\Lambda\to\infty$.

\paragraph{The scale anomaly and the beta function:} 
The quartic interaction preserves scale invariance at tree level in
$d = 4$, but the one-loop correction breaks it. We now show that the
parameter space Ward identity captures this breaking directly: the
Ward violating term $\mathcal{B}[\mathcal{O}_4]$ is nonzero at
$O(\lambda^2)$, and its connected amputated projection gives the
one-loop beta function. 

Interacting $\phi^4$ theory is defined in NN-FT~\cite{demirtas2023neural}
by deforming the parameter density\footnote{For mathematical ease, we do not normal order $\phi^4$ in the density deformation.
This permits tadpole self-contractions that generate one-particle
reducible contributions to the connected four-point function,
including power-divergent mass corrections of order
$\lambda\Lambda^2$. These do not enter the one-particle irreducible
vertex and therefore do not affect the logarithmic running or the
beta function extracted below.}
\begin{equation}
    p_\lambda(\theta)
    = p_G(\theta)\,\exp\!\Big[
      -\frac{\lambda}{4!}\int d^4 x\;\phi_\theta^4(x)\Big]
    =: p_G(\theta)\,e^{-V_4(\theta)}\,,
\end{equation}
with $p_G$ the free theory density. The Ward violating
term~\eqref{eq:B0-def} generalizes to
\begin{equation}
\label{eq:break_insertion}
    \mathcal{B}[\mathcal{O}]
    = -\langle B(\theta)\,\mathcal{O}\rangle_\lambda
      + F_\lambda[\mathcal{O}]\,,
\end{equation}
where the expectation and boundary flux $F_\lambda$ are now computed
with respect to $p_\lambda$. Rewriting interacting expectations as
$\langle \mathcal{O} \rangle_\lambda
= \langle \mathcal{O}\,e^{-V_4}\rangle_0 /
\langle e^{-V_4}\rangle_0$
and expanding in powers of $\lambda$, we obtain
$\mathcal{B} = \sum_k \mathcal{B}^{(k)}$. To demonstrate that scale
invariance is broken, it suffices to show that
$\mathcal{B}[\mathcal{O}] \neq 0$ at some order for some $\mathcal{O}$. 

We focus on the case $\mathcal{O} = \mathcal{O}_4$, specifically the
connected amputated projection of $\mathcal{B}^{(2)}[\mathcal{O}_4]$ (henceforth the $[\mathcal{O}_4]$ is implied), since it
isolates the momentum dependent part of the one-loop four-point
function and directly yields the beta function.
After the cumulant expansion
(Appendix~\ref{app:scale_cumulant}), the connected amputated
projection reduces to $\mathcal{B}^{(2),\text{c.a.}} =
N_2^{\text{c.a.}}$, where $N_2$ involves the composite observable
$X = \frac{1}{2}V_4^2\,\mathcal{O}_4$. We evaluate the bulk and
flux contributions in turn; the detailed calculation is in
Appendix~\ref{app:scale_bulk_flux}, and we summarize the logic here.

We first wish to see how the one-loop four-point function arises.
For the bulk, $X$ is homogeneous of degree $12$ in the Gaussian
coefficients $a_j$, since $V_4^2$ contributes degree $8$ and
$\mathcal{O}_4$ contributes degree $4$. Gaussian integration by
parts converts $\frac{1}{\sigma^2}\sum_j \langle a_j^2\,X\rangle_0$
into $(N + 12)\langle X\rangle_0$. Together with
$B = -\frac{2}{\sigma^2}\sum_j a_j^2 + 6N$, this gives
\begin{equation}
    -\langle B\,X\rangle_0^{\text{conn}}
    = (24 - 4N)\,\Delta^{(2),\text{conn}}(x_i)\,,
\end{equation}
where $\Delta^{(2),\text{conn}}(x_i)
:= \langle X\rangle_0^{\text{conn}}$ is the connected one-loop
four-point function in position space.
For the flux, expanding $D_b X$ via the product rule and using
$D_b V_4 = 8V_4$ produces another multiple of
$\Delta^{(2),\text{conn}}$ plus a spacetime dilatation piece:
\begin{equation}
    F_0[X]^{\text{conn}}
    = (4N - 20)\,\Delta^{(2),\text{conn}}(x_i)
      + \left\langle \tfrac{1}{2}V_4^2
        \sum_{r=1}^4 x_r^\mu \partial_{x_r^\mu} \mathcal{O}_4
        \right\rangle_0^{\text{conn}}\,.
\end{equation}
Adding the bulk and flux we have
\begin{equation}
\label{eq:B2-position}
    \mathcal{B}^{(2),\text{conn}}(x_i)
    = \left(4 + \sum_{r=1}^4 x_r^\mu \partial_{x_r^\mu}\right)
      \Delta^{(2),\text{conn}}(x_i)\,.
\end{equation}
since the extensive $O(N)$ terms cancel
identically.

We see the Ward violating term is the position space dilatation operator
acting on the one-loop four-point function. This has the same
structure as the free theory result~\eqref{eq:B0-result}, but with a
crucial difference: the one-loop four-point function has logarithmic
momentum dependence, which persists as
$\Lambda \to \infty$, violating the Ward identity.

We now extract the beta function. Passing to momentum space and
amputating (Appendix~\ref{app:B1_mom}), the constant pieces in the
dilatation operator cancel against the scaling of the external
propagators and the momentum conserving delta function, leaving
\begin{equation}
\label{eq:B2_amp}
    \mathcal{B}^{(2),\text{c.a.}}(p_i)
    = -\sum_{r=1}^4 p_r^\mu
      \frac{\partial}{\partial p_r^\mu}\Delta^{(2)}(p_i)\,,
\end{equation}
where $\Delta^{(2)}(p_i)$ is the amputated one-loop vertex.
Evaluating at the Euclidean symmetric point
$p_i^2 = \mu^2\ll\Lambda^2$, $p_i \cdot p_j = -\mu^2/3$ ($i \neq j$), so that
$s = t = u = 4\mu^2/3$, gives (Appendix~\ref{app:loop_integral})
\begin{equation}
    \Delta^{(2)}(\mu)
    = \frac{3\lambda^2}{2}
      \int_{|k| \leq \Lambda}
      \frac{d^4k}{(2\pi)^4}\,
      \frac{1}{k^2(k - q)^2}
    = \frac{3\lambda^2}{32\pi^2}\,
      \log\frac{\Lambda^2}{\mu^2}
    + \text{finite}\,,
\end{equation}
with $q^2 = 4\mu^2/3$ and the cutoff inherited from the compact
support of the frequency density. The factor of $3$ counts the three
channel pairings ($s$, $t$, $u$), each giving the same integral at
the symmetric point. Along the symmetric ray where,
$\mathcal{B}^{(2),\text{c.a.}}(\mu) = -\mu\,d\Delta^{(2)}/d\mu$.
Recalling that at one-loop in $\phi^4$ theory in $4d$, the renormalized coupling is
\begin{equation}
    \lambda_R(\mu) := \lambda - \Delta^{(2)}(\mu)
    + O(\lambda^3)
\end{equation}
and the beta function by
$\beta(\mu) := \mu\,d\lambda_R/d\mu$, we see that
\begin{equation}
    \beta = \mathcal{B}^{(2),\text{c.a.}}(\mu)\;,
\end{equation}
the one-loop beta function equals the Ward violating term at the
symmetric point. From~\eqref{eq:B2_amp}
\begin{equation}
\label{eq:phi4-beta-result}
\boxed{
    \beta
    = \frac{3\lambda^2}{16\pi^2}
      + O\!\left(\frac{\mu^2}{\Lambda^2}\right),
}\;.
\end{equation}
the standard one-loop result for the $\beta$-function of $\phi^4$ theory.

The novelty of our calculation is that this result (relying on~\eqref{eq:B2_amp})
is derived entirely from the
parameter space Ward identity. The logarithmic scale dependence of
the loop integral is the genuine anomaly: unlike the power suppressed
cutoff artifacts in the free theory, it persists as
$\Lambda \to \infty$ and signals the breaking of scale invariance by
quantum effects.

\subsection{Weyl anomaly, bosonic string, and critical dimension}
\label{sec:weyl}

We now derive the critical dimension $D = 26$ of the bosonic string by computing the Weyl anomaly in the NN-FT framework.
We place the matter and ghost architectures on $S^2$ and study the response of the partition function to a constant Weyl rescaling $g_{ab} \to e^{2\epsilon} g_{ab}$.
We use the sphere as the worldsheet because after conformal gauge fixing on $S^2$, the $c$ and $\bar c$ ghost fields each have three zero modes corresponding to the conformal Killing vectors (CKVs).
For any ghost number zero operator $\mathcal{O}$, we define the reduced sphere correlator
\begin{equation}
\label{eq:primed-sphere-correlator}
\vev{\mathcal{O}}'_{S^2} :=
\vev{c(z_1)c(z_2)c(z_3)\,
\bar{c}(\bar{z}_1)\bar{c}(\bar{z}_2)\bar{c}(\bar{z}_3)\,
\mathcal{O}}_{S^2}\,,
\end{equation}
with the matter zero mode volume factored out.
(Without the ghost dressing, $\mathcal{O}$ has a trivial one-point function on $S^2$.)
The reduced correlator accounts for the residual $PSL(2,\mathbb{C})$ gauge symmetry and enables us to compute the Weyl anomaly~\cite{Friedan:1985ge}.

\paragraph{Architectures on $S^2$:}
Following~\cite{FerkoHalversonMutchler2026}, we replace flat space random Fourier features
with eigenfunction expansions on the sphere.
Let $\hat{n} \in S^2 \subset \mathbb{R}^3$.
The matter architecture is
\begin{equation}
\label{eq:X-sphere}
X^\mu(\hat{n}) = \sum_{\ell=0}^{L} \sum_{m=-\ell}^{\ell}
a^\mu_{\ell,m}\, Y_{\ell m}(\hat{n})\,,
\end{equation}
where $Y_{\ell m}$ are real spherical harmonics.
We specify the theory by giving the architecture together with the unnormalized measure $d\mu = u(\theta)\,d\theta$ on parameter space.
The normalized density $p(\theta) = u(\theta)/\mathcal{N}$ determines all correlators as usual; however, the unnormalized partition function $Z_u = \int d\theta\; u(\theta)$ retains additional metric dependent information that will be essential for the anomaly, as we explain below.

For the scalar nonzero modes ($\ell \geq 1$), the unnormalized density is
\begin{equation}
\label{eq:scalar-unnorm}
u^{(X)}(a) = \prod_{\ell=1}^{L}\prod_{m=-\ell}^{\ell}\prod_{\mu=1}^{D}
\exp\!\left(-\frac{(a^\mu_{\ell,m})^2}{2\sigma_\ell^2}\right)\,,
\qquad
\sigma_\ell^2 = \frac{2\pi\alpha'}{\ell(\ell+1)}\,,
\end{equation}
independently in $\mu$, $\ell$, $m$.
The scalar zero mode $a^\mu_{0,0}$ carries a flat density; taking $\sigma_0 \to \infty$ enforces spacetime translation invariance.
The two-point function is
\begin{equation}
\label{eq:XX-sphere}
\langle X^\mu(\hat{n}_1)\, X^\nu(\hat{n}_2) \rangle
= \delta^{\mu\nu}\biggl[\sigma_0^2
+ \frac{\alpha'}{2} \sum_{\ell=1}^{L}
\frac{2\ell+1}{\ell(\ell+1)}\, P_\ell(\cos\gamma)\biggr]\,,
\end{equation}
where $\cos\gamma = \hat{n}_1 \cdot \hat{n}_2$.
The $\ell$-sum reproduces $-\frac{\alpha'}{2}\log\!\bigl(\frac{1-\cos\gamma}{2}\bigr) + \text{const}$ as $L \to \infty$ (Appendix~\ref{app:sphere_two_pt}), recovering the standard flat space propagator in stereographic coordinates.

The ghost architectures use spin-weighted spherical harmonics ${}_s Y_{\ell m}$~\cite{NewmanPenrose1966,Goldberg1967}, which exist only for $\ell \geq |s|$.
The holomorphic ghosts $b \equiv b_{zz}$ (spin-weight $s = +2$) and $c \equiv c^z$ (spin-weight $s = -1$) have the architecture
\begin{equation}
\label{eq:bc-sphere}
b(\hat{n}) = \sum_{\ell=2}^{L} \sum_{m=-\ell}^{\ell}
\beta_{\ell m}\; {}_{-2} Y_{\ell m}^*(\hat{n})\,, \qquad
c(\hat{n}) = \sum_{\ell=1}^{L} \sum_{m=-\ell}^{\ell}
\chi_{\ell m}\; {}_{-1} Y_{\ell m}(\hat{n})\,.
\end{equation}
The constraint $\ell \geq |s|$ forces the $b$-sum to start at $\ell = 2$ while the $c$-sum starts at $\ell = 1$: the three $\ell = 1$ modes of $c$ are the CKVs of $S^2$.
The unnormalized Grassmann Gaussian density~\cite{FrankHalversonMaitiRuehle2025, FrankHalverson2026}
for the paired modes ($\ell \geq 2$) is
\begin{equation}
\label{eq:ghost-density}
u^{(b,c)}(\beta,\chi) = \prod_{\ell=2}^{L}\prod_{m=-\ell}^{\ell}
\exp\!\left(\frac{\mu_\ell}{4\pi}\, \beta_{\ell m}\, \chi_{\ell m}\right)\,,
\qquad
\mu_\ell = \sqrt{(\ell{-}1)(\ell{+}2)}\,,
\end{equation}
so that $\mathbb{E}[\beta_{\ell m}\chi_{\ell' m'}] = (4\pi/\mu_\ell)\,\delta_{\ell\ell'}\delta_{mm'}$.
The eigenvalues $\mu_\ell$ are those of the spin-lowering operator $\bar\eth$ acting on spin-weight-$(-1)$ harmonics.
The three unpaired CKV parameters $\chi_{1,m}$ carry a flat density ($u = 1$), the Grassmann analogue of the scalar zero mode prescription; their Berezin integrals vanish unless saturated by the ghost insertions $c(z_1)c(z_2)c(z_3)$ in the reduced correlator~\eqref{eq:primed-sphere-correlator}.
In the $L \to \infty$ limit, the two-point function reproduces the standard ghost OPE $\langle b_{zz}(z)\, c^w(w)\rangle = 1/(z-w)$ in stereographic coordinates (Appendix~\ref{app:ghost_prop}).
The antiholomorphic ghosts $\bar b \equiv \bar b_{\bar z\bar z}$ and $\bar c \equiv \bar c^{\bar z}$ are an independent copy with spin-weights $-2$ and $+1$:
\begin{equation}
\label{eq:bc-bar-sphere}
\bar{b}(\hat{n}) = \sum_{\ell=2}^{L} \sum_{m=-\ell}^{\ell}
\bar\beta_{\ell m}\; {}_{2} Y_{\ell m}^*(\hat{n})\,, \qquad
\bar{c}(\hat{n}) = \sum_{\ell=1}^{L} \sum_{m=-\ell}^{\ell}
\bar\chi_{\ell m}\; {}_{1} Y_{\ell m}(\hat{n})\,,
\end{equation}
with the same density structure and all cross-correlators between chiralities vanishing.

\paragraph{Weyl anomaly from the parameter space measure:}
In order to specify a theory, we must start by defining where the theory lives, for example on $\mathbb{R}^4$.
In the examples of Sections~\ref{sec:ml_examples} and~\ref{sec:com_scalar}, the architecture is determined at the beginning once and for all using this specification, and the network does not thereafter refer explicitly to the background.
Changing parameters moves through a fixed space of field configurations.
For the case at hand, the situation is different.
The Weyl transformation rescales the metric by sending 
\begin{equation}
    g \to e^{2\epsilon}g \qquad \sigma_\ell^2 \to e^{2\epsilon} \sigma_\ell^2,
\end{equation}
We are working on a different sphere as a consequence of the transformation, and therefore the architecture of the NN-FT changes.
The absorption mechanism relates the new architecture to the old one via a parameter redefinition.
The theory remains Gaussian, but the eigenvalue dependent variances in the density also shift.
A Weyl transformation therefore takes us from one NN-FT to a genuinely different one, and the question is how the partition function responds.
It should be stressed that changing the architecture and changing the density are not competing descriptions; they are two gauges/trivializations of the same background dependent NN-FT.

The normalized generating functional of the theory can be written in the form
\begin{align}\label{ZJg_defn}
    Z [ J, g ] = \frac{Z_u [ J, g ] }{Z_u [ 0, g ] } \, .
\end{align}
Here the subscript $u$ indicates that the corresponding quantities are expectation values taken with respect to an unnormalized measure $u ( \theta_g )$, for instance
\begin{align}
    Z_u [ J, g ] = \int \dd\theta_g \,  u(\theta_g)\; \exp\big( {\int_{S^2_g} J \phi} \big) 
\end{align}
where the subscripts emphasize the metric dependence.
Although operator expectation values can be obtained by taking functional derivatives of $Z_u [ J, g ]$ and then normalizing by dividing by the factor $Z_u [ 0, g ]$ as in~\eqref{ZJg_defn}, we stress that this normalization factor itself has physical meaning, even in the conventional field space approach to QFT.
For instance, in finite temperature quantum field theory, $Z_u [ 0, g ] = \mathrm{Tr} ( e^{- \beta H} )$ is the thermal partition function which determines the free energy.
In real time QFT, $Z_u [ 0, g ]$ can be interpreted as a vacuum-to-vacuum amplitude which captures information about vacuum bubble diagrams.

A similar phenomenon is observed in the NN-FT setting.
In previous sections, we have defined a NN-FT via an architecture $\phi_\theta$ along with a \emph{normalized} density $p ( \theta )$ over its parameters.
However, in the present setting, we wish to study the Weyl anomaly, which is encoded by the change in the normalization factor $Z_u [ 0, g ]$ in response to a change of the metric. Via direct computation of the part of $Z_u[0,g]$ arising from free boson network weights, we have 
\begin{equation}
    Z_u[0,g]\big |_\text{free boson}= \prod_{\ell \geq 1, m, \mu} \sqrt{2\pi \sigma_\ell^2}
\;\propto\; (\det{}'\Delta)^{-D/2} \, .
\end{equation}
where the proportionality arises from the relationship between the variance and Laplacian eigenvalues. This is the functional determinant whose Weyl variation yields the Weyl anomaly, which we will compute directly in NN-FT. Note that our NN-FT architecture realizes the conventional free boson whose action is, after integration by parts, $S = \frac{1}{4 \pi \alpha'} \int_{S^2} \sqrt{g} X^\mu \Delta X_\mu$. After expanding $X^\mu$ in an orthonormal basis of the Laplacian, we therefore see that each mode contributes its associated Laplacian eigenvalue to the action, which sets the mode's Euclidean energy cost in the weight $e^{-S}$.

Let us specify in detail how the free boson and ghost NN-FTs change under a constant Weyl rescaling. The Laplacian eigenvalues scale as $\ell(\ell{+}1) \to e^{-2\epsilon}\ell(\ell{+}1)$, giving $\sigma_\ell^2 \to e^{2\epsilon}\sigma_\ell^2$, while the
spin-lowering eigenvalues scale as $\mu_\ell \to e^{-\epsilon}\mu_\ell$.
The orthonormal eigenfunctions on the rescaled background are $\widetilde{f}_n = e^{-(1-t)\epsilon}f_n$, where $t = n_{\text{lower}} - n_{\text{upper}}$ counts net lower indices (Appendix~\ref{app:weyl_scaling}).
Absorbing this rescaling into the coefficients,
\begin{equation}\label{eq:weyl-generators}
\xi^{a^\mu_{\ell m}} = -a^\mu_{\ell m}\,,\qquad
\xi^{\beta_{\ell m}} = +\beta_{\ell m}\,,\qquad
\xi^{\chi_{\ell m}} = -2\chi_{\ell m}\,,
\end{equation}
for scalars ($t{=}0$), $b$-ghosts ($t{=}+2$), and $c$-ghosts ($t{=}-1$) respectively, with the same generators for the antiholomorphic ghost sector.

We now compute the Weyl variation of the unnormalized partition function $Z_u = \int d\theta\; u_g(\theta)$ sector by sector, showing that in each case the score piece of the breaking function cancels and only the Jacobian piece survives.
Recall that a change of variables $\theta \to \theta'$ transforms the measure as $d\theta = (1 - \epsilon\,B_{\text{Jac}})\,d\theta'$, where
\begin{equation}\label{eq:BJac-sign}
B_{\text{Jac}} =
\begin{cases}
+\sum_a \partial_a\xi^a & \text{(bosonic)}\,,\\
-\sum_a \partial_a\xi^a & \text{(Grassmann)}\,,
\end{cases}
\end{equation}
the sign difference reflecting the inverse determinant in the Berezin measure.

\medskip\noindent\emph{Scalar nonzero modes:} 
Consider a single mode with
$u_g(a) = e^{-a^2/(2\sigma_\ell^2)}$ and
$u_{\widetilde{g}}(a) = e^{-a^2/(2e^{2\epsilon}\sigma_\ell^2)}$.
The absorption gives $a = (1{+}\epsilon)a'$, so
$da = (1 - \epsilon\,B_{\text{Jac}})\,da'$
with $B_{\text{Jac}} = \partial_a\xi^a = -1$, and
\begin{equation}\label{eq:scalar-weyl-cov}
Z_u[0,{\widetilde{g}}]
= (1 - \epsilon\,B_{\text{Jac}})
  \int da'\;
  \exp\!\left(
    -\frac{(1+\epsilon)^2\, a'^2}
          {2\,e^{2\epsilon}\sigma_\ell^2}
  \right).
\end{equation}
In the exponent,
$(1+\epsilon)^2 e^{-2\epsilon} = 1 + O(\epsilon^2)$:
the reabsorption into primed parameters and the Weyl rescaled variance produce equal and opposite $O(\epsilon)$ shifts that cancel, corresponding to the score piece of the breaking function being compensated by the Weyl variation of the density.
The Jacobian, on the other hand, remains:
\begin{equation}
Z_u[0,{\widetilde{g}}] = (1 - \epsilon\,B_{\text{Jac}})\,Z_u[0,g]\,,
\qquad
\delta_\epsilon\log Z_u[0,g] = -\epsilon\,B_{\text{Jac}}\,.
\end{equation}
\medskip\noindent\emph{Ghost paired modes:}
The calculation has the same structure.
For a single $(\beta_{\ell m}, \chi_{\ell m})$ pair at
$\ell \geq 2$ with one chirality, the unnormalized density is
$u_g = e^{(\mu_\ell/4\pi)\beta\chi}$.
On the rescaled background,
$\mu_\ell \to e^{-\epsilon}\mu_\ell$, so
$u_{\widetilde{g}} = e^{(e^{-\epsilon}\mu_\ell/4\pi)\beta\chi}$.
The absorption gives
$\beta = (1{-}\epsilon)\beta'$ and
$\chi = (1{+}2\epsilon)\chi'$.
In the exponent,
$e^{-\epsilon}(1{-}\epsilon)(1{+}2\epsilon) = 1 + O(\epsilon^2)$,
so the score piece cancels and the exponent returns to
$(\mu_\ell/4\pi)\beta'\chi'$.
The Berezin measure transforms as
$d\beta\,d\chi
= (1 - \epsilon\,B_{\text{Jac}})\,d\beta'\,d\chi'$
with
$B_{\text{Jac}}
= -(\partial_\beta\xi^\beta + \partial_\chi\xi^\chi)
= -(1-2) = +1$.
Therefore
\begin{equation}
Z_u[0,{\widetilde{g}}] = (1 - \epsilon\,B_{\text{Jac}})\,Z_u[0,g]\,,
\qquad
\delta_\epsilon\log Z_u[0,g] = -\epsilon\,B_{\text{Jac}}
\end{equation}
per pair per chirality.

\medskip\noindent\emph{CKV and zero modes:}
The three unpaired CKV parameters $\chi_{1,m}$ have flat
density $u = 1$ and generator $\xi^\chi = -2\chi$.
The exponent is trivial, and the Berezin measure gives
$B_{\text{Jac}} = -\partial_\chi\xi^\chi = +2$ per mode,
so $\delta_\epsilon\log Z_u[0,g] = -\epsilon\,B_{\text{Jac}}$
per CKV per chirality.
The scalar zero mode $a^\mu_{0,0}$ has flat density and
generator $\xi^a = -a$, giving
$B_{\text{Jac}} = \partial_a\xi^a = -1$ and
$\delta_\epsilon\log Z_u[0,g] = -\epsilon\,B_{\text{Jac}}$
per component.

In every sector the result is the same: the score piece of the breaking function cancels against the
Weyl variation of the eigenvalue spectrum, and
\begin{equation}
\delta_\epsilon\log Z_u[0,g] = -\epsilon\,B_{\text{Jac}}\,.
\end{equation}
This is the finite-dimensional analogue of the Fujikawa anomaly.
Since each generator is linear in its own parameter, $B_{\text{Jac}}$ is a $\theta$-independent constant in each sector, making the total anomaly a pure mode count:
\begin{equation}
B_\text{Jac}^{(X)} = -D\biggl[\sum_{\ell=1}^{L}(2\ell{+}1) + 1\biggr]\,,
\qquad B_\text{Jac}^{(\text{ghost})} = 2\sum_{\ell=2}^L(2\ell{+}1) + 12\,.
\end{equation}
We regularize the divergent sums using spectral zeta functions
\begin{equation}
\zeta_X(s) = \sum_{\ell=1}^{\infty}(2\ell{+}1)[\ell(\ell{+}1)]^{-s}\,,
\qquad
\zeta_{bc}(s) = \sum_{\ell=2}^{\infty}(2\ell{+}1)\mu_\ell^{-s}\,,
\end{equation}
evaluated at $s = 0$ in Appendix~\ref{app:zeta}:
$\zeta_X(0) = -2/3$ and
$\zeta_{bc}(0) = -5/3$.
The regularized total is
\begin{align}
B_\text{anom}^\text{reg}
&= -D\bigl[\zeta_X(0) + 1\bigr]
   + 2\,\zeta_{bc}(0) + 12
\notag\\
&= -\frac{D}{3} - \frac{10}{3} + 12
\;=\; -\frac{1}{3}(D-26)\,.
\label{eq:total-weyl-anomaly}
\end{align}
Thus, demanding $\delta_\epsilon\log Z_u[0,g] = 0$ requires
\begin{equation}
    \boxed{D = 26}\,.
\end{equation}
This reproduces the critical dimension of the bosonic string.\footnote{
To be precise, when the total central charge of the $X^\mu$ and the $bc$ system vanishes, the Weyl transformation can be treated as a true redundancy of the matter plus ghost system.
When $c_\text{tot}\neq 0$, the Weyl factor becomes the Liouville mode rather than a pure gauge degree of freedom.
In the latter case, we have a non-critical string theory.
}
In string theory, Weyl invariance of the worldsheet partition function is required for consistent gauge fixing of the worldsheet metric in the Polyakov path integral; the NN-FT calculation recovers the correct anomaly coefficient from a regularized $L \to \infty$ limit of a finite mode count in parameter space.

\subsection{Winding and higher-form symmetries}
\label{sec:higher-form}

The goal of this section is to illustrate that higher-form symmetries and their Ward identities are most naturally studied in NN-FT using a \emph{mixed} parameter space that has been enriched to include additional topological labels, some of which are typically discrete.

This is in contrast with the examples in the preceding sections, which exclusively involved transformations generated by smooth vector fields $\xi^a(\theta)$ on a continuous parameter space.
In that case the Ward identity arises from the vanishing of the right side of the contracted Schwinger--Dyson equation
\begin{equation}
\bigl\langle \delta_\xi \mathcal O\bigr\rangle = -\bigl\langle B_\xi\,\mathcal O\bigr\rangle \,, 
\qquad B_\xi=\xi\cdot s+\partial\cdot\xi \, ,
\end{equation}
up to possible boundary fluxes.

Compact theories introduce a different kind of parameter space structure in which the NN-FT ensemble contains both conventional neural network parameters and topological sector data.
In such cases, we write the expectation value schematically as
\begin{equation}
\label{eq:hf-mixed-expectation}
\langle \mathcal{O} \rangle = \sum_Q \int d\theta\; P( \theta, Q )\,\mathcal{O} [ \phi_{\theta, Q} ] \, ,
\end{equation}
as in~\cite{Ferko:2026ken}.
Here, $\theta$ denotes the continuous parameters in a fixed topological sector.
Such sectors are labeled by the topological data $Q$, which generically includes both discrete random variables (winding labels, vortex charges, etc.) and continuous quantities associated with topological sectors (\textit{e.g.}, the position of a vortex or defect).
If such continuous topological parameters are present, the sum over those $Q$ in (\ref{eq:hf-mixed-expectation}) is replaced by an integral.

Schwinger--Dyson identities hold in each continuous sector, for fixed values of the discrete labels.
Let us write the sector data as $(Q,\zeta)$, where $Q$ denotes discrete topological labels and $\zeta$ denotes continuous defect moduli such as vortex positions.
Then
\begin{equation}
\langle\mathcal O\rangle = \sum_Q\int_{\mathcal M_Q} d\theta\,d\zeta\; P_q(\theta,\zeta)\,\mathcal O[\phi_{\theta,q,\zeta}] \,.
\end{equation}
For a smooth vector field $\Xi=\Xi^A\partial_A$ acting in $\mathcal M_Q$, with $A$ running over both ordinary neural parameters and continuous sector coordinates, the contracted Schwinger--Dyson identity gives
\begin{equation}
\big\langle \Xi^A\partial_A\mathcal O\big\rangle = -\big\langle B_{\Xi,q}\mathcal O\big\rangle+\mathcal F_\Xi[\mathcal O] \,,
\qquad B_{\Xi,q} = \Xi^A\partial_A\log P_q+\partial_A\Xi^A \,,
\end{equation}
and $\mathcal F$ is the boundary flux when the continuous sector has boundary.

For simplicity, we momentarily restrict to the case where all of the topological parameters $Q$ are discrete.\footnote{
The higher-form winding symmetry we consider in Appendix~\ref{app:bkt} is not such a case, however; it is represented by a topological operator acting diagonally on the sector decomposition.}
In the next part of this section, we will consider the winding symmetry of a compact scalar, which offers an example of a symmetry that is not generated by such a flow on a fixed-$Q$ sector.
This is to be expected; parameter space vector fields such as $\xi^a ( \theta )$ are defined on smooth manifolds, and cannot carry components along a parameter direction $Q$ which is inherently discrete rather than smooth.
Rather, the Ward identity associated with such a winding symmetry is instead a topological sector identity.

\paragraph{The compact boson and its shift symmetry:}
Consider a compact boson $\chi$ in $d$ spacetime dimensions with normalized periodicity
\begin{equation}
\chi \sim \chi + 2\pi\,.
\end{equation}
The shift symmetry $\chi\mapsto\chi+\lambda$ is still of the type studied in Sections~\ref{sec:sd-main} and~\ref{sec:ward}.
If the compact zero mode $\chi_0\in S^1$ is included among the continuous parameters, then the shift is absorbed by
\begin{equation}
\xi^{\chi_0}=\lambda\,,
\qquad
\xi^a=0\quad\text{for all other parameters.}
\end{equation}
For a translation invariant density on the zero mode circle, $\partial_{\chi_0}\log P=0$, and since $\partial_{\chi_0}\xi^{\chi_0}=0$,
\begin{equation}
B_\text{shift}=0\,.
\end{equation}
Thus, the ``electric'' shift symmetry is represented by the same divergence-free parameter space current condition as before.
Its charged operators are the vertex operators $e^{in\chi}$.

\paragraph{The winding symmetry:}
The compact scalar has another symmetry, the topological winding symmetry.
One can view this as a ``magnetic'' symmetry, which is dual to the preceding ``electric'' shift symmetry in a sense analogous to electric-magnetic duality.
The conventional $(d-1)$-form current associated to the $(d-2)$-form winding symmetry is
\begin{align}
J_w^{(d-1)} = \frac{1}{2 \pi} \ast \dd \chi \, .
\end{align}
Equivalently, its Hodge dual one-form is
\begin{align}
j_w^{(1)} = \ast J_w^{(d-1)} = \frac{1}{2 \pi} \dd \chi \, .
\end{align}
The topological symmetry operator is naturally written as an integral of $j_w^{(1)}$ over a closed one-cycle.
For smooth configurations,
\begin{equation}
\label{eq:hf-bianchi-new}
\dd j_\text{w}^{(1)}=0\,,
\end{equation}
which is a Bianchi identity rather than an equation of motion.
Since $\chi$ is circle valued, $\dd\chi$ should be understood globally as a closed one-form with integral periods, even when no single valued real lift of $\chi$ exists.

This current should not be confused with the parameter space current $j^a=\xi^a p$ introduced in Section~\ref{sec:sd-main}.
The chain described in~\eqref{eq:chain} applies to Ward identities generated by smooth parameter space flows.
The winding current is instead determined directly by the compact field configuration, and in the NN-FT representation by the discrete sector label $Q$.
We emphasize that, although one could attempt to describe this winding symmetry using the techniques of Section~\ref{sec:sd-main} and Section~\ref{sec:ward} by engineering a NN-FT architecture and parameter density for the field dual to the scalar $\chi$, this is not possible if one insists upon working with a NN-FT realization of the compact scalar $\chi$ itself.

The charge measured on a closed one-cycle $\gamma$ is
\begin{equation}
\label{eq:hf-winding-charge-new}
W(\gamma) = \int_\gamma j_\text{w}^{(1)} = \frac{1}{2\pi}\oint_\gamma \dd\chi \in \mathbb Z\,.
\end{equation}
The corresponding symmetry operator is
\begin{equation}
\label{eq:hf-winding-operator-new}
U_\alpha(\gamma) = \exp\!\left(i\alpha\int_\gamma j_\text{w}^{(1)}\right)
= \exp\!\left( \frac{i\alpha}{2\pi}\oint_\gamma \dd\chi \right) \,,
\qquad \alpha\sim\alpha+2\pi\,.
\end{equation}
This is the structure of a $q=d-2$ form global symmetry:
the symmetry operators are supported on closed one-cycles, and the charged objects are $(d{-}2)$-dimensional defects~\cite{Gaiotto:2014kfa}. Although we now turn to the specific example of vortex defects and the NN-FT realization of their higher-form symmetries, we refer the reader to~\cite{Sharpe:2015mja,Gomes:2023ahz,Schafer-Nameki:2023jdn,Brennan:2023mmt,Bhardwaj:2023kri,Shao:2023gho} for pedagogical introductions to various aspects of generalized global symmetries.

\paragraph{Vortex sectors and the linking NN-FT Ward identity:}
A vortex defect of charge $m\in\mathbb Z$, supported on an oriented codimension two submanifold $\Sigma^{d-2}$, is charged under $U(1)_\text{w}$. In the presence of such a defect the Bianchi identity is modified to account for the source localized on $\Sigma$,
\begin{equation}
\label{eq:hf-vortex-source-single-new}
\dd j_\text{w}^{(1)} = m\,\delta^{(2)}_\Sigma\,,
\end{equation}
where $\delta^{(2)}_\Sigma$ is the two-form Poincar\'e dual to $\Sigma$.
We fix orientation conventions by
\begin{equation}
\label{eq:hf-pd-normalization-new}
\int_D \delta^{(2)}_\Sigma = \operatorname{Int}(D,\Sigma) = \operatorname{Link}(\partial D,\Sigma) \,,
\end{equation}
for any oriented two-chain $D$ whose boundary is disjoint from $\Sigma$.
Consider a fixed topological sector of the full NN-FT ensemble which contains a collection of defects $Q$ supported on submanifolds $\Sigma_a$ and with charges $m_a$.
In this setting one has
\begin{equation}
\label{eq:hf-vortex-source-many-new}
\dd j_\text{w}^{(1)} = \sum_a m_a\,\delta^{(2)}_{\Sigma_a}\,.
\end{equation}
On a space with non-trivial topology, there are also ``harmonic'' winding sectors.
For instance, a NN-FT on a torus $T^d$ admits configurations in which line operators wrap the one-cycles of the torus.
Neglecting such harmonic winding sectors, if $\gamma = \partial D$, then Stokes' theorem yields
\begin{equation}
\label{eq:hf-linking-charge-new}
\int_\gamma j_\text{w}^{(1)} = \sum_a m_a\,\operatorname{Link}(\gamma,\Sigma_a)\,.
\end{equation}
The winding symmetry is therefore represented in parameter space by the sector observable
\begin{equation}
U_\alpha(\gamma;\theta,Q) = e^{i\alpha W_Q(\gamma)} \,.
\end{equation}
Its insertion in a correlator is
\begin{equation}
\big\langle U_\alpha(\gamma)\,\mathcal O\big\rangle = \sum_Q\int d\theta\; P(\theta,Q)\, e^{i\alpha W_Q(\gamma)} \mathcal O[\phi_{\theta,Q}] \,.
\end{equation}
Thus $U_\alpha(\gamma)$ acts diagonally on the sector decomposition of the parameter space integral, i.e. it multiplies each sector contribution by the phase determined by its winding charge.
It does not move $(\theta,Q)$ to a neighboring point in parameter space, as in the preceding examples.

The linking Ward identity follows directly from the distributional Bianchi identity.
A vortex defect $V_m(\Sigma)$ imposes the boundary condition
\begin{equation}
\oint_{S^1_{\rm link}} d\chi = 2\pi m
\end{equation}
on a small circle linking $\Sigma$.
Equivalently,
\begin{equation}
\dd j_w^{(1)} = m\,\delta^{(2)}_\Sigma \,,
\qquad j_w^{(1)}=\frac{1}{2\pi} \dd\chi \,.
\end{equation}
If $\gamma=\partial D$ and $D$ is chosen transverse to $\Sigma$, then Stokes' theorem gives
\begin{equation}
\int_\gamma j_w^{(1)} = \int_D \dd j_w^{(1)} = m\int_D\delta^{(2)}_\Sigma = m\,\operatorname{Link}(\gamma,\Sigma) \,.
\end{equation}
Therefore,
\begin{equation}
U_\alpha(\gamma) = \exp\!\left(i\alpha\int_\gamma j_w^{(1)}\right)
\end{equation}
acts on a vortex insertion by the phase
\begin{equation}
\exp\!\left(i\alpha m\,\operatorname{Link}(\gamma,\Sigma)\right) \,.
\end{equation}

Thus, for external vortex defects,
\begin{equation}
\left\langle
U_\alpha(\gamma)\prod_r V_{m_r}(\Sigma_r)\,\mathcal O \right\rangle = \exp\!\left( i\alpha\sum_r m_r\,\operatorname{Link}(\gamma,\Sigma_r) \right) \left\langle \prod_r V_{m_r}(\Sigma_r)\,\mathcal O \right\rangle
\end{equation}
whenever $U_\alpha(\gamma)$ can otherwise be removed or deformed away from $\mathcal O$, and ignoring harmonic winding sectors.

A selection rule follows only when the symmetry operator can be compared with the identity, for example when it can be moved to infinity or to a trivial cycle in the relevant correlator.
In that case invariance for all $\alpha$ requires the total charge detected by that removable cycle to
vanish.
Loops enclosing only a subset of charged insertions generally give Ward phases, not vanishing conditions as in Section~\ref{sec:ward}.
In the NN-FT ensemble this is implemented as a statement about the sector observable $W_Q(\gamma)$.
The insertion $U_\alpha(\gamma)$ does not differentiate fields or move continuous parameters; it multiplies the $(\theta,Q)$ sector contribution by $e^{i\alpha W_Q(\gamma)}$.
Thus, the Ward identity is not a consequence of a smooth parameter space vector field.
It is a topological identity of the sector decomposition
\begin{equation}
\sum_Q\int d\theta\,P(\theta,Q)\, e^{i\alpha W_Q(\gamma)}\mathcal O[\phi_{\theta,Q}] \,.
\end{equation}
Selection rules arise only when the topological operator can be deformed to a trivial cycle, in which case exactness for all $\alpha$ requires the total charge detected by that removable cycle to vanish.

\paragraph{Exactness and breaking by the sector measure:}
For smooth absorbed symmetries, unbroken symmetry is the continuity equation~\eqref{eq:continuity}, or equivalently $B=0$.
For the winding higher-form symmetry, the corresponding condition is instead a support condition on the
topological sector measure.
In the absence of external charged defects, an exact continuous $U(1)_\text{w}$ symmetry requires the vacuum ensemble to be supported on source free sectors,
\begin{equation}
\label{eq:hf-source-free-support-new}
\dd j_\text{w}^{(1)}[\chi_{\theta,Q}]=0 \, ,
\end{equation}
apart from the possible harmonic winding sectors mentioned above.
This is the higher-form analogue of~\eqref{eq:continuity}.

If one attempts to remove a contractible symmetry loop $\gamma$, the failure is measured by
\begin{equation}
\label{eq:hf-topological-breaking-insertion-new}
\Delta_\alpha(\gamma;\theta,Q) := U_\alpha(\gamma)[\theta,Q]-1 = e^{i\alpha W_Q(\gamma)}-1\,.
\end{equation}
Infinitesimally,
\begin{equation}
\Delta_\alpha(\gamma;\theta,Q) = i\alpha\,W_Q(\gamma)+O(\alpha^2)\,.
\end{equation}
The quantity $\Delta_\alpha$ is not the breaking function $B_\xi$ of Sections~\ref{sec:sd-main} and~\ref{sec:ward}, because no smooth vector field $\xi^a$ generates the winding symmetry on the original compact scalar parameter space.
Rather, $\Delta_\alpha$ is the topological Ward defect: it measures the obstruction to removing a contractible symmetry operator from a given sector.
This also clarifies the distinction between external and dynamical vortices.
External vortex defects are charged insertions.
They do not break the symmetry; they produce the linking/contact terms in the Ward identity.
Dynamical vortex defects are different.
If the vacuum sector measure sums over charged vortex configurations with nonzero fugacity, then contractible loops can detect charges in the vacuum ensemble.
The continuous winding symmetry is then explicitly broken, or reduced to a subgroup if only a sublattice of charges is dynamical.\footnote{
This phenomenon is familiar from standard treatments of higher-form symmetries.
For instance, the inclusion of charged matter in Maxwell theory breaks the electric $U(1)$ one-form global symmetry which the theory enjoys in the absence of charged matter.}
This is analogous to density dependent explicit breaking in $B=\xi\cdot s+\partial \cdot\xi$, but the breaking lives in the topological sector distribution rather than in a smooth score function.
In this parameterization there is no finite-dimensional Jacobian term $\sum_a\partial_a\xi^a$ associated with the winding symmetry.

A two-dimensional example is the NN-FT realization of the Berezinskii--Kosterlitz--Thouless (BKT) transition~\cite{Ferko:2026ken}, where the topological data $Q$ are encoded in the positions $x_a$ and charges $m_a$ of the vortices.
We give further details about this example in Appendix~\ref{app:bkt}.

\paragraph{Higher-dimensional extension:}
The same structure extends to higher dimensions.
In $d=3$, vortex defects are closed loops $\mathcal C_a$, and a closed one-cycle $\gamma$ measures their linking:
\begin{equation}
W_Q(\gamma) = W_\text{harm}(\gamma;Q) + \sum_a m_a\,\operatorname{Link}(\gamma,\mathcal C_a)\,.
\end{equation}
In $d=4$, vortices are closed surfaces, and in general $d$ they are closed $(d{-}2)$-dimensional submanifolds.
Closure,
\begin{equation}
\partial\Sigma_a=0\,,
\end{equation}
expresses the absence of endpoints or monopole events.
If endpoints are allowed, then $\dd\delta^{(2)}_{\Sigma_a}$ is the Poincar\'e dual of $\partial\Sigma_a$, up to the orientation convention, and the endpoint defects explicitly violate the higher-form symmetry.

On a compact background, the total vortex source must also obey the cohomological consistency condition
\begin{equation}
\label{eq:hf-cohomological-neutrality-new}
\left[ \sum_a m_a\,\delta^{(2)}_{\Sigma_a} \right] = 0 \in H^2(M,\mathbb Z) \,,
\end{equation}
whenever it is to be written globally as $\dd j_\text{w}^{(1)}$.
In two dimensions this reduces to the familiar global neutrality condition $\sum_a m_a=0$.
In higher dimensions, it is a condition on the total homology class of the vortex worldvolumes.

The lesson for NN-FT is that higher-form Ward identities are naturally implemented in a mixed parameter space.
Smooth continuous parameters still obey the contracted Schwinger--Dyson identities of Section~\ref{sec:sd-main}; the higher-form symmetry is encoded by the topological sector observable $U_\alpha(\gamma)[Q]$, and unbroken symmetry is a support condition on the sector measure.
This is the topological analogue of the condition $B=0$, but it is not itself a statement about the divergence of a smooth parameter space current.

\subsection{T-duality 
}\label{sec:T}
The higher-form discussion above shows that compact NN-FTs naturally have mixed parameter spaces: ordinary continuous neural network parameters are supplemented by discrete topological labels.
For the compact coordinate of a closed bosonic string, these labels are the momentum and winding numbers.
T-duality is therefore a particularly clean NN-FT example of a finite symmetry/duality whose breaking is measured by a Radon--Nikodym ratio rather than by the infinitesimal function $B_\xi=\xi\cdot s+\partial\cdot\xi$ on one smooth sheet.
We focus on one compact target coordinate $X\sim X+2\pi R$ and suppress spectator noncompact coordinates and ghosts, which are inert in the discussion below.

\paragraph{NN-FT architecture:}
We use a doubled zero mode bookkeeping for the compact coordinate.
This does not introduce an independent local field; it packages the ordinary zero mode and the dual zero mode in a way that makes both momentum and winding selection rules visible in parameter space.
Let
\begin{equation}
\label{eq:tdual-nnft-parameter-space-new}
\Theta_R=(\theta_L,\theta_R,x_0,\widetilde x_0;n,w)
\in \mathcal M_R
:=\mathcal M_{\rm osc}\times S^1_R\times S^1_{\widetilde R}\times\mathbb Z^2 \,,
\qquad
\widetilde R=\frac{\alpha'}{R} \,.
\end{equation}
Here, $x_0\sim x_0+2\pi R$, $\widetilde x_0\sim \widetilde x_0+2\pi\widetilde R$, and $(n,w)\in\mathbb Z^2$ label momentum and winding sectors.
The NN-FT expectation value is
\begin{equation}
\label{eq:tdual-hf-mixed-ensemble}
\langle\mathcal O\rangle_R = \sum_{n,w\in\mathbb Z}\int d\mu_{\rm osc}(\theta_L,\theta_R) \frac{dx_0}{2\pi R}\frac{d\widetilde x_0}{2\pi\widetilde R}\, P_R(n,w)\,\mathcal O\!\left[X^{(R)}_{\Theta_R}\right] \,.
\end{equation}
For simplicity, we take the positive lattice weight with vanishing real modular parameter,
\begin{equation}
\label{eq:tdual-nnft-lattice-weight-new}
P_R(n,w)=\frac{1}{Z_{\rm lat}(R)} \exp\!\left[-\pi\tau_2\left(\alpha'\frac{n^2}{R^2}+\frac{R^2}{\alpha'}w^2\right)\right] \,.
\end{equation}
Although we are studying fields on a cylinder, this is the same weight that appears in the torus partition function for vanishing real modular parameter $\tau_1 = 0$.
The normalization satisfies $Z_{\rm lat}(R)=Z_{\rm lat}(\widetilde R)$ after exchanging $n$ and $w$.
Equivalently,
\begin{equation}
\label{eq:tdual-nnft-weight-identity-new}
P_R(n,w)=P_{\widetilde R}(w,n) \,.
\end{equation}

In chiral coordinates $u=\tau+\sigma$ and $v=\tau-\sigma$, the compact coordinate is represented as
\begin{align}
X(\tau,\sigma) &=X_L(u)+X_R(v) ~, \notag \\
X_L ( u ) &= \frac12(x_0+\widetilde x_0) +\frac{\alpha'}{2}p_L(n,w)\,u +X_L^{\rm osc}(u) \,, \label{eq:lrmovers} \\
X_R ( v ) &= \frac12(x_0-\widetilde x_0) +\frac{\alpha'}{2}p_R(n,w)\,v +X_R^{\rm osc}(v) \,, \notag
\end{align}
where we have defined
\begin{equation}\label{eq:momenta}
p_L(n,w)=\frac{n}{R}+\frac{wR}{\alpha'}\,,
\qquad p_R(n,w)=\frac{n}{R}-\frac{wR}{\alpha'}\,.
\end{equation}
The zero modes $x_0$ and $\widetilde{x}_0$ are sampled from uniform distributions on $[0, 2 \pi R)$ and $[0, 2 \pi \widetilde{R})$, respectively, while we represent the oscillator term using the architecture
\begin{equation}
X_{L/R}^{\rm osc}(\xi)=\sqrt{\frac{\alpha'}{2}}\sum_{m=1}^{M}
\frac{a_m^{L/R}\cos(m\xi)+b_m^{L/R}\sin(m\xi)}{\sqrt m} \, .
\label{eq:compact_chiral_features}
\end{equation}

The T-duality map is the finite transformation
\begin{equation}
\label{eq:tdual-hf-map}
\mathsf T_R: (R;n,w) \longmapsto \left(\widetilde R;w,n\right) \,,
\qquad \widetilde R=\frac{\alpha'}{R} \,.
\end{equation}
At the level of chiral fields this is accompanied by
\begin{equation}
X_L\longmapsto X_L \,,
\qquad X_R\longmapsto -X_R \,,
\end{equation}
so that $p_L$ is fixed and $p_R$ changes sign.

This is a finite transformation between two mixed NN-FT measures, not an infinitesimal flow.
To discuss such finite transformations, it is convenient to package the full parameter density over continuous and discrete parameters into a single symbol.
In the language of measure theory, the mixed parameter density~\eqref{eq:tdual-hf-mixed-ensemble} defines a measure $\mu_R$ on the measurable space of parameters $\Theta_R$.
The full parameter space can be viewed as a disjoint union of ``sheets''
\begin{equation}
\mathcal M_R=\bigsqcup_{Q\in\mathbb Z^2}\mathcal M_{R,Q} \,,
\end{equation}
where each sheet $\mathcal M_{R,Q}$ carries an ordinary probability distribution over the continuous parameters.
Locally, on each sheet, one may write $d\mu_R(\vartheta,Q)= \rho_R(\vartheta,Q)\,d \vartheta $, where $\vartheta = ( x_0, \widetilde{x}_0, \theta )$.
We assume $d\mu_{\rm osc}$ is invariant under right-moving parity $(a_m^R,b_m^R)\mapsto-(a_m^R,b_m^R)$.

\paragraph{Continuous compact symmetries:}
Before discussing T-duality, it is useful to identify the ordinary breaking functions for the two compact zero mode shifts.
The momentum $U(1)$ shift acts by
\begin{equation}
\label{eq:tdual-nnft-momentum-shift-new}
x_0\mapsto x_0+\epsilon R \,,
\qquad \xi_x^{x_0}=R \,,
\end{equation}
with all other parameters fixed.
Because the density in~\eqref{eq:tdual-hf-mixed-ensemble} is uniform on $S^1_R$,
\begin{equation}
\label{eq:tdual-nnft-momentum-breaking-new}
B_x=\xi_x^{x_0}\partial_{x_0}\log\rho_R + \partial_{x_0}\xi_x^{x_0}=0 \,.
\end{equation}
Similarly, the winding $U(1)$ shift acts on the dual zero mode by
\begin{equation}
\label{eq:tdual-nnft-winding-shift-new}
\widetilde x_0\mapsto \widetilde x_0+\epsilon \widetilde R \,,
\qquad
\xi_{\widetilde x}^{\widetilde x_0}=\widetilde R \,,
\end{equation}
and its breaking function also vanishes,
\begin{equation}
\label{eq:tdual-nnft-winding-breaking-new}
B_{\widetilde x}=\xi_{\widetilde x}^{\widetilde x_0}\partial_{\widetilde x_0}\log\rho_R+\partial_{\widetilde x_0}\xi_{\widetilde x}^{\widetilde x_0}=0 \,.
\end{equation}
Thus these two compact global symmetries fit directly into the infinitesimal NN-FT formalism of Sections~\ref{sec:sd-main} and~\ref{sec:ward}.
For the vertex operators defined using~\eqref{eq:lrmovers} and~\eqref{eq:momenta},
\begin{equation}
\label{eq:tdual-nnft-vertex-new}
V_{n,w}^{(R)}(z,\bar z)=\ :\exp\!\left(i p_L(n,w)X_L(z)+i p_R(n,w)X_R(\bar z)\right): \,,
\end{equation}
the two zero mode shifts give
\begin{equation}
\delta_x V_{n,w}^{(R)}=i\epsilon n\,V_{n,w}^{(R)} \,,
\qquad \delta_{\widetilde x} V_{n,w}^{(R)}=i\epsilon w\,V_{n,w}^{(R)} \,.
\end{equation}
Since $B_x=B_{\widetilde x}=0$, the Ward identities imply the usual compact boson selection rules
\begin{equation}
\label{eq:tdual-nnft-selection-rules-new}
\left(\sum_i n_i\right) \left\langle\prod_i V_{n_i,w_i}^{(R)}\right\rangle_R=0 \,,
\qquad \left(\sum_i w_i\right) \left\langle\prod_i V_{n_i,w_i}^{(R)}\right\rangle_R=0 \,.
\end{equation}
In a presentation that does not include the dual zero mode, the second identity appears instead as the topological sector Ward identity discussed in Section~\ref{sec:higher-form}.

\paragraph{Finite breaking functions:}
T-duality is not generated by a smooth vector field on a fixed-radius sheet.
It is a finite bijection between two members of the NN-FT family, the theory at radius $R$ and the theory at radius $\widetilde R=\alpha'/R$.
For any such bijection
\begin{equation}
\mathsf T:\mathcal M_R\longrightarrow \mathcal M_{\widetilde R} \,,
\end{equation}
define the finite NN-FT Jacobian ratio\footnote{
We think of $(\mathsf T_R)_*\mu_R$ as the pushforward of the mixed measure from~\eqref{eq:tdual-hf-mixed-ensemble} to the dual radius parameter space.
The Radon--Nikodym theorem states that, under the assumption that one measure must be absolutely continuous with respect to the other, which is always satisfied in cases of our interest, any pair of measures on the same measurable space can be related to one another via a ratio known as the Radon--Nikodym derivative.}
and finite breaking function by
\begin{equation}
\label{eq:tdual-nnft-finite-breaking-def-new}
\mathcal J_{\mathsf T_R}(\Theta') :=\frac{d(\mathsf T_*\mu_R)}{d\mu_{\widetilde R}}(\Theta') \,,
\qquad \mathfrak B_{\mathsf T_R}:=1-\mathcal J_{\mathsf T_R} \,.
\end{equation}
Then, for any observable on the $\widetilde R$ theory,
\begin{equation}
\label{eq:tdual-nnft-finite-ward-new}
\boxed{
\langle \mathcal O\circ\mathsf T\rangle_R-
\langle\mathcal O\rangle_{\widetilde R}
=-\langle\mathfrak B_{\mathsf T}\,\mathcal O\rangle_{\widetilde R} \,.}
\end{equation}
This is the finite analogue of
$\langle\delta_\xi\mathcal O\rangle=-\langle B_\xi\mathcal O\rangle$.
We should compare this to the continuous case: when $\mathsf g_\epsilon$ is the infinitesimal flow of a smooth parameter space vector field $\xi$ parametrized by $\epsilon$, one has
\begin{equation}\label{eq:tdual-hf-RN}
\mathcal J_{\mathsf g_\epsilon}=1-\epsilon(\xi\cdot s+\partial\cdot\xi)+O(\epsilon^2) \,,
\qquad \mathfrak B_{\mathsf g_\epsilon}=1-\mathcal J_{\mathsf g_\epsilon} = \epsilon B_\xi+O(\epsilon^2) \,.
\end{equation}
The finite breaking function in~\eqref{eq:tdual-nnft-finite-breaking-def-new} packages the same two ingredients as the infinitesimal one: a density ratio, which is the finite version of the score term, and an ordinary Jacobian on the continuous sheets.
For purely discrete relabelings the latter is replaced by the counting measure Jacobian, which is one for a bijection.

The T-duality map is
\begin{equation}
\label{eq:tdual-nnft-map-new}
\mathsf T_R: (\theta_L,\theta_R,x_0,\widetilde x_0;n,w;R)\longmapsto (\theta_L,-\theta_R,\widetilde x_0,x_0;w,n;\widetilde R) \,,
\end{equation}
where $-\theta_R$ denotes the right-moving oscillator parity in~\eqref{eq:compact_chiral_features}.
At the level of chiral fields,
\begin{equation}
\label{eq:tdual-nnft-chiral-action-new}
X_L\longmapsto X_L \,,
\qquad X_R\longmapsto -X_R \,,
\end{equation}
because
\begin{equation}
\label{eq:tdual-nnft-momenta-transform-new}
p_L^{(\widetilde R)}(w,n)=p_L^{(R)}(n,w) \,,
\qquad p_R^{(\widetilde R)}(w,n)=-p_R^{(R)}(n,w) \,.
\end{equation}
The oscillator measure is invariant under $\theta_R\mapsto-\theta_R$; the zero mode measure
\begin{equation}
\frac{dx_0}{2\pi R}\frac{d\widetilde x_0}{2\pi\widetilde R}
\end{equation}
is mapped to the corresponding dual radius zero mode measure; and the sector density obeys~\eqref{eq:tdual-nnft-weight-identity-new}.
Therefore,
\begin{equation}
\label{eq:tdual-nnft-Bzero-new}
\mathcal J_{\mathsf T}=1 \,,
\qquad
\mathfrak B_{\mathsf T}=0 \,,
\end{equation}
and the finite Ward identity becomes
\begin{equation}
\label{eq:tdual-nnft-exact-ward-new}
\boxed{
\langle\mathcal O\circ\mathsf T_R\rangle_R = \langle\mathcal O\rangle_{\widetilde R} \,.}
\end{equation}
This is the NN-FT statement of T-duality: it is a measure preserving relabeling of a mixed parameter space, together with a transformation of the hyperparameter $R$.
It is not a consequence of differentiating with respect to $R$; the radius labels different NN-FT ensembles.

Equivalently, for compact boson vertex operators
\begin{equation}
V_{n,w}^{(R)}(z,\bar z) = :\!\exp\!\left( i p_L X_L(z)+i p_R X_R(\bar z) \right)\!: \,,
\end{equation}
the T-duality map gives
\begin{equation}
V_{n,w}^{(R)} \longleftrightarrow V_{w,n}^{(\widetilde R)} \,.
\end{equation}
Therefore the operator level ``Ward identity'' is\footnote{We refer to~\eqref{eq:tdual-hf-vertex-ward} as a Ward identity by analogy with the discussion in Section~\ref{sec:ward}, but we note that this equation is not really a constraint within a single theory, but a relation between correlators in different theories at radii $R$ and $\widetilde{R}$.
As we will see shortly, at the self-dual radius it is a genuine Ward identity.}
\begin{equation}
\label{eq:tdual-hf-vertex-ward}
\boxed{ \left\langle \prod_{i=1}^N V_{n_i,w_i}^{(R)}(z_i,\bar z_i) \right\rangle_R
= \left\langle \prod_{i=1}^N V_{w_i,n_i}^{(\widetilde R)}(z_i,\bar z_i) \right\rangle_{\widetilde R} \,,}
\end{equation}
with the usual compact zero mode neutrality conditions imposed on either side.
The equality is not derived from differentiating with respect to a continuous parameter.
It is an identity of the discrete momentum--winding sector measure.

\paragraph{Fixed radius and the self-dual point:}
It is useful to contrast this exact duality with the momentum--winding exchange
\begin{equation}
\mathsf S_R:(n,w)\mapsto(w,n) \,,
\end{equation}
performed \emph{without} transforming the radius.
For $R\neq\sqrt{\alpha'}$ this is not an absorbed symmetry of the full compact coordinate NN-FT: the target zero mode circle and the dual zero mode circle have different radii, and the lattice density is not invariant.
As a diagnostic on the discrete sector measure, its finite breaking function, from~\eqref{eq:tdual-nnft-lattice-weight-new}, is
\begin{equation}
\label{eq:tdual-nnft-fixed-radius-breaking-new}
\mathfrak B_{\mathsf S_R}(n,w) = 1-\frac{P_R(w,n)}{P_R(n,w)}
= 1-\exp\!\left[-\pi\tau_2\left(\frac{\alpha'}{R^2}-\frac{R^2}{\alpha'}\right)(w^2-n^2)\right] \,.
\end{equation}
This vanishes for every lattice sector if and only if
\begin{equation}
\label{eq:tdual-nnft-self-dual-radius-new}
R=\sqrt{\alpha'} \,.
\end{equation}
Thus, away from the self-dual point, momentum--winding exchange is not a symmetry of one fixed radius ensemble; it is an equivalence between two different ensembles.
At the self-dual point, $\mathcal M_R=\mathcal M_{\widetilde R}$ and $\mathsf T_R$ becomes a genuine $\mathbb Z_2$ symmetry of a single NN-FT measure:
\begin{equation}
\label{eq:tdual-nnft-self-dual-ward-new}
\langle\mathcal O\circ\mathsf S-\mathcal O\rangle_{R=\sqrt{\alpha'}}=0.
\end{equation}
This is the finite/discrete analogue of the condition $B_\xi=0$ for a smooth absorbed symmetry.
The statement above detects the T-duality $\mathbb Z_2$ itself.
The familiar enhanced current algebra at the self-dual radius requires also including the extra dimension one vertex operators in the set of symmetry generators; their Ward identities are additional current algebra Ward identities, not merely the finite lattice relabeling in~\eqref{eq:tdual-nnft-self-dual-ward-new}. However, it is also possible to characterize this enhanced $SU(2)_L \times SU(2)_R$ symmetry which emerges at the self-dual radius using the machinery of NN-FT and breaking functions. This example is explored in Appendix~\ref{sec:su2}.

\section{Conclusions}
\label{sec:conclusions}

In this work, we have further developed the proposed correspondence between quantum field theories and ensembles of random neural networks~\cite{Halverson:2020trp,Halverson:2021aot}, focusing on the role of symmetries and anomalies.
We have presented a parameter space analogue of the Schwinger--Dyson equations, which are traditionally obtained using a function space treatment of the partition function, using only the fundamental theorem of calculus and conventional (Riemann or Lebesgue) integrals over neural network parameters.
After specializing to field variations which represent putative symmetries of a given NN-FT, we obtained Ward identities which are satisfied when a particular breaking function
\begin{align}
    B = \sum_a \xi^a s_a + \sum_a \partial_a \xi^a
\end{align}
vanishes, where $s_a:= \partial_a \log{p}$ is the \emph{score}.
We have argued that, in special cases, our parameter space conserved currents reproduce familiar local currents in quantum field theory; however, it appears that the formalism developed here is more general, and can potentially be used to describe conserved quantities in field theories which admit no local Lagrangian description.
We have also illustrated this formalism in several examples which are of interest both in artificial intelligence --- including multilayer perceptrons and transformers --- and in quantum field theory, such as in the toy example of a complex scalar with a $U(1)$ symmetry.

A classical symmetry which is broken by quantum effects is said to exhibit an (ABJ-type) anomaly. Using our parameter space machinery and the language of breaking functions, we have studied two such anomalies.
The scale anomaly of a $4d$ scalar field theory with a $\phi^4$ interaction can be understood from the parameter space perspective, including a quantitative calculation of the one-loop beta function.
Likewise, using an appropriate architecture for the scalars and $bc$ ghosts, one can study the Weyl anomaly of bosonic string theory and reproduce the critical dimension $D = 26$ from an NN-FT calculation.

Finally, we have explored the properties of symmetries (and their associated Ward identities) in NN-FTs which exhibit non-trivial topological features.
The topological winding symmetry of a compact boson illustrates an example which does not immediately admit a natural description in terms of our treatment of Ward identities involving continuous parameter space flows, but which can be incorporated into the NN-FT paradigm after enlarging the parameter space to include discrete topological labels, following the philosophy of~\cite{Ferko:2026ken}.
Dual descriptions of the same quantum field theory, such as the equivalence of bosonic string theory on target spacetimes involving circles of radius $R$ and $\widetilde{R} = \alpha'/R$, can also be studied using this framework, and we have used the breaking function perspective to reproduce the fact that string theory with a compact target exhibits an enhanced $\mathbb{Z}_2$ symmetry at the self-dual radius.

There remain several promising directions for future research.
Since we have considered ABJ-type anomalies, a natural question, of course, is to see the chiral anomaly~\cite{Adler1969,BellJackiw1969} in a Dirac NN-FT.
In a separate, forthcoming paper, we carry out this analysis within the fermionic NN-FT architecture~\cite{Frank:2025zuk}: a finite width Grassmann network defines the spinor field through its neural features and  global $U(1)_V$ and massless $U(1)_A$ act as explicit linear transformations of the neural network parameters.
When vector and axial background fields are introduced as neural network source deformations, the anomaly is obtained from studying the divergence of a regulated axial vector field on the coefficient space.
As in the Fujikawa construction~\cite{Fujikawa1979}, the regulator plays a critical role, but in this case, it is constructed natively in the NN-FT parameter space. 

Discrete symmetries, such as parity, can also be anomalous.
There is no infinitesimal breaking function: the diagnostic is in the finite transformation law of the measure.
For a discrete symmetry element $h\in G$, the NN-FT presentation over a background $b$ may be carried to the presentation over a transformed background $h\cdot b$, with parameters related by a finite change of variables
\begin{equation}
A_h:\Theta_b\to \Theta_{h\cdot b} \,,
\qquad
\Phi_{h\cdot b}(A_h\theta)=R_h\Phi_b(\theta) \,,
\end{equation}
where $R_h$ is the induced action on fields.
As in the case of a Weyl transfomation, an apparent change in architecture can be absorbed into a relabeling of parameters, or equivalently traded for a change of density in a fixed trivialization.
The possible anomaly is the residual finite Jacobian or phase in the unnormalized weight:
\begin{equation}
\rho_{h\cdot b}(A_h\theta)\, \left|\operatorname{Ber} D A_h(\theta)\right|
= e^{i\alpha_h(b)}\rho_b(\theta) \,.
\end{equation}
If the phase $\alpha_h(b)$ can be removed by a local counterterm or by changing the trivialization of the background dependent density, then the symmetry is non-anomalous.
If not, the symmetry acts only projectively on the partition function and correlation functions.
Consistency under group multiplication requires the phases to obey the cocycle condition
\begin{equation}
\alpha_{h_1h_2}(b) = \alpha_{h_1}(h_2\!\cdot b)+\alpha_{h_2}(b) \mod 2\pi \,,
\end{equation}
which tells us that applying $h_2$ and then $h_1$ must give the same anomalous phase as applying the combined group element $h_1h_2$.
A removable anomaly corresponds to a cocycle that can be eliminated by redefining the phase of the partition function.
In general, discrete anomalies appear in the transformation of the normalization constant, in signs or phases of determinants and Pfaffians, in twisted sectors, and in the projective transformation law of observables.

This anticipates a treatment of 't Hooft anomalies~\cite{tHooft:1979rat}.
In addition to the ABJ-type anomalies we have studied in this work (classical symmetries which are broken by quantum effects), it would be useful to understand  through the lens of NN-FT what happens when global symmetries cannot be gauged.
This is especially interesting because, by virtue of the Wess--Zumino consistency condition, 't Hooft anomalies are topological and therefore invariant along renormalization group (RG) flows.
Thus if one can compute such 't Hooft anomalies at one point along an RG trajectory, one can then extrapolate and make inferences about the behavior of the infrared theory (for instance, to determine whether it is gapless, trivially gapped, or flows to a TQFT in the IR).

We already observe something like a 't Hooft anomaly in our discussion of higher-form symmetries.
To probe an anomalous symmetry here using NN-FT, we must introduce a background $(q+1)$-form gauge field $\mathcal B$ coupled to the higher-form current and study background gauge transformations
\begin{equation}
\mathcal B\mapsto \mathcal B+\dd\Lambda \,.
\end{equation}
For a mixed NN-FT measure $d\nu_{\mathcal B}(\theta,Q)$, the analogue of the breaking function is the infinitesimal Radon--Nikodym/Jacobian variation
\begin{equation}
B_\Lambda^{\rm NN}(\theta,Q;\mathcal B) = \left.\frac{d}{dt}\right|_{t=0} \log\frac{ \dd\nu_{\mathcal B+t\,\dd\Lambda}(g_t\cdot(\theta,Q))}{d\nu_{\mathcal B}(\theta,Q)} + \left.\frac{d}{dt}\right|_{t=0}\log J_{g_t} \,.
\end{equation}
An anomaly is present when this variation cannot be removed by local counterterms and instead produces a background dependent phase.

Another future direction is to conduct a more systematic investigation of generalized global symmetries using NN-FT.
Since NN-FT primarily offers a representation of correlation functions of local operators, we have argued that a neural network treatment of higher-form symmetries associated with extended operators requires the inclusion of additional parameters which label the number and configuration of such extended operators.
It would be very interesting to make this proposal more precise and to study other examples, such as the $\mathbb{Z}_N$ $1$-form center symmetry of $SU(N)$ gauge theory, in neural network language.
Likewise, one would like to give a NN-FT characterization of higher-group and non-invertible global symmetries; one potentially promising target for the latter might be the non-invertible Kramers--Wannier defect line $\mathcal{D}$ of the $2d$ Ising model.
As we have mentioned, gauging higher-form symmetries requires the introduction of background tensor fields with two or more indices, which represents another direction for future research.

Finally, although the spirit of the present work is AI-for-physics, it would be intriguing to investigate whether any of our results have implications for the opposite direction of physics-for-AI.
For instance, the characterization of symmetry preserving densities via the continuity equation~\eqref{eq:continuity} could inform density design in machine learning applications.

We aim to return these questions in future work.

\begin{acknowledgments}
We thank Fabian Ruehle for discussions.
We acknowledge the open-source Get Physics Done (GPD) project by Physical Superintelligence PBC (PSI), whose AI-assisted physics research workflow was helpful in carrying out aspects of this work.
We also acknowledge the language models Claude Opus 4.6, GPT-5.4, and GPT-5.5.
The authors take full responsibility for the accuracy of this work, having reviewed, verified, and approved all AI-generated content. 
C.F.\ and J.H.\ are supported by the National Science Foundation under Cooperative Agreement PHY-2019786 (the NSF AI Institute for Artificial Intelligence and Fundamental Interactions).
J.H. is also supported by NSF grant PHY-220990.
V.J.\ is supported by the South African Research Chairs Initiative of the Department of Science, Technology, and Innovation and the National Research Foundation (grant 78554).
\end{acknowledgments}

\appendix

\addtocontents{toc}{\protect\setcounter{tocdepth}{1}}

\section{Scale anomaly: computational details}
\label{app:scale_details}

This appendix collects the calculations underlying the scale anomaly
derivation in \S\ref{sec:scale-anomaly}. We work throughout with the
free theory expectation $\langle\cdot\rangle_0$, the breaking
function $B = -\frac{2}{\sigma^2}\sum_j a_j^2 + 6N$, and the
boundary flux~\eqref{eq:flux}.

\subsection{Perturbative expansion and cumulant structure}
\label{app:scale_cumulant}

Rewriting interacting expectations as free theory expectations
weighted by $e^{-V_4}$, the Ward violating
term~\eqref{eq:break_insertion} becomes
\begin{equation}
\label{eq:B_combined_app}
    \mathcal{B}[\mathcal{O}_4]
    = \frac{-\langle B\,\mathcal{O}_4\,e^{-V_4}\rangle_0
            + F_0[\mathcal{O}_4\,e^{-V_4}]}
           {\langle e^{-V_4}\rangle_0}\,.
\end{equation}
Write the numerator as a power series $\sum_k N_k$ with terms
\begin{equation}
\label{eq:Nk-def-app}
    N_k := \frac{(-1)^k}{k!}
    \bigl(-\langle B\,\mathcal{O}_4\,V_4^k\rangle_0
    + F_0[\mathcal{O}_4\,V_4^k]\bigr)\,,
\end{equation}
and the denominator as
$\langle e^{-V_4}\rangle_0 = \sum_k D_k$ with
$D_k = (-1)^k \langle V_4^k\rangle_0 / k!$.
Inverting order by order, the $O(\lambda^2)$ contribution is
\begin{equation}
\label{eq:cumulant_app}
    \mathcal{B}^{(2)}
    = N_2 - D_1\,N_1 + (D_1^2 - D_2)\,N_0\,.
\end{equation}
Restricting to the connected sector, each $D_k$ is a
momentum independent vacuum number, so multiplication by $D_k$ does
not alter which contraction patterns appear:
\begin{equation}
\label{eq:B2conn-expanded-app}
    \mathcal{B}^{(2),\text{conn}}
    = N_2^{\text{conn}}
    - D_1 \cdot N_1^{\text{conn}}
    + (D_1^2 - D_2) \cdot N_0^{\text{conn}}\,.
\end{equation}

The result~\eqref{eq:B0-result} established that $N_0$ is the
dilatation of the free four-point function
$\langle\mathcal{O}_4\rangle_0$, which at $N \to \infty$ is purely
disconnected (a sum of products of two-point functions). Since the
dilatation acts by the product rule on individual contractions and
preserves connectivity, $N_0^{\text{conn}} = 0$.

Setting $Y = V_4\,\mathcal{O}_4$, the observable $Y$ is homogeneous
of degree $8$ in the Gaussian amplitudes $a_j$. The same
bulk plus flux calculation described in
Appendix~\ref{app:scale_bulk_flux} (with degree $8$ replacing $12$)
gives
\begin{equation}
    N_1^{\text{conn}}(x_i)
    = \left(4 + \sum_{r=1}^4 x_r^\mu \partial_{x_r^\mu}\right)
      \Delta^{(1),\text{conn}}(x_i)\,,
\end{equation}
where $\Delta^{(1),\text{conn}}$ is the connected tree-level
four-point function. The tree-level amputated vertex
$\Delta^{(1)}$ is a momentum independent constant
($\propto \lambda$), so after Fourier transformation and amputation
(Appendix~\ref{app:B1_mom}), $N_1^{\text{c.a.}}
= -\sum_r p_r^\mu \partial_{p_r^\mu} \Delta^{(1)} = 0$.

Therefore, it suffices to evaluate
$N_2^{\text{conn}}$.

\subsection{Bulk and flux evaluation at $O(\lambda^2)$}
\label{app:scale_bulk_flux}

Define $X = \frac{1}{2}V_4^2\,\mathcal{O}_4$ and
$\Delta^{(2),\text{conn}}(x_i)
:= \langle X \rangle_0^{\text{conn}}$, the connected one-loop
four-point function in position space.

The observable $X$ is homogeneous of degree $12$ in the Gaussian
amplitudes $a_j$ ($V_4^2$ contributes degree $8$, $\mathcal{O}_4$
contributes degree $4$), so Euler's theorem gives
$\sum_j a_j\,\partial_{a_j} X = 12\,X$.
Using Gaussian integration by parts,
$\langle a_j^2\,f \rangle_0 = \sigma^2 \langle f \rangle_0
+ \sigma^2 \langle a_j\,\partial_{a_j} f \rangle_0$,
and summing over $j$:
\begin{equation}
    \frac{1}{\sigma^2} \sum_{j=1}^N
    \langle a_j^2\,X \rangle_0
    = N \langle X \rangle_0
    + \sum_{j=1}^N
      \langle a_j\,\partial_{a_j} X \rangle_0
    = (N + 12)\langle X \rangle_0\,.
\end{equation}
Hence, using $B = -\frac{2}{\sigma^2}\sum_j a_j^2 + 6N$,
\begin{equation}
\label{eq:bulk-N2-app}
    -\langle B\,X \rangle_0^{\text{conn}}
    = (24 - 4N)\,\Delta^{(2),\text{conn}}(x_i)\,.
\end{equation}

Expanding the derivative in the flux~\eqref{eq:flux}:
\begin{align}
F_0[X]
&= \frac{1}{Z_0} \sum_{j,\mu}
   \int d\theta\;\Bigl[
   \underbrace{(\partial_{b_{j\mu}} b_{j\mu})}_{=\,1}\,
   p\,X
   + b_{j\mu}\,(\partial_{b_{j\mu}} p)\,X
   + b_{j\mu}\,p\,\partial_{b_{j\mu}} X
   \Bigr]
   \notag\\
&= 4N\,\langle X \rangle_0
   - \langle D_b\,X \rangle_0\,,
\end{align}
where the score term $\partial_{b_{j\mu}} p$ vanishes in the
interior of $B^4_\Lambda$ because the frequency density is uniform,
and $D_b := -\sum_{j,\mu} b_{j\mu}\,\partial_{b_{j\mu}}$ is the
frequency space dilatation operator.

Using the product rule,
$D_b X = \frac{1}{2}(D_b V_4^2)\,\mathcal{O}_4
       + \frac{1}{2}V_4^2\,(D_b \mathcal{O}_4)$.
Classical marginality of $V_4$ in $d = 4$ gives
$D_b V_4 = 8V_4$ and hence $D_b V_4^2 = 16 V_4^2$.
For the external fields,
$D_b \mathcal{O}_4
= (4 - \sum_{r=1}^4 x_r^\mu \partial_{x_r^\mu})\mathcal{O}_4$.
Projecting onto the connected sector:
\begin{equation}
\label{eq:flux-N2-app}
    F_0[X]^{\text{conn}}
    = (4N - 20)\,\Delta^{(2),\text{conn}}(x_i)
      + \left\langle \tfrac{1}{2}V_4^2
        \sum_{r=1}^4 x_r^\mu \partial_{x_r^\mu} \mathcal{O}_4
        \right\rangle_0^{\text{conn}}\,.
\end{equation}

Adding~\eqref{eq:bulk-N2-app} and~\eqref{eq:flux-N2-app}, the
extensive $O(N)$ terms cancel:
\begin{equation}
\label{eq:B2-position-app}
    N_2^{\text{conn}}(x_i)
    = \left(4 + \sum_{r=1}^4 x_r^\mu \partial_{x_r^\mu}\right)
      \Delta^{(2),\text{conn}}(x_i)\,,
\end{equation}
which is~\eqref{eq:B2-position} in the main text.

\subsection{Amputated momentum-space dilatation operator}
\label{app:B1_mom}

We show that the position-space dilatation
\begin{equation}
\label{eq:B1_x_short_app}
    \mathcal{B}(x_i)
    = \left(4 + \sum_{r=1}^4 x_r^\mu \partial_{x_r^\mu}\right)
      \Delta(x_i)
\end{equation}
reduces, after Fourier transformation and amputation, to
\begin{equation}
\label{eq:B1_target_app}
    \mathcal{B}^{\text{c.a.}}(p_i)
    = -\sum_{r=1}^4 p_r^\mu
      \frac{\partial}{\partial p_r^\mu} \Delta(p_i)\,,
\end{equation}
where $\Delta(p_i)$ denotes the amputated vertex in either the
$O(\lambda)$ or $O(\lambda^2)$ case. The argument is the same
for both; we suppress the superscript.

Define the Fourier transforms
\begin{equation}
    \Delta(p_i)
    := \int \prod_{r=1}^4 d^4 x_r\;
       e^{-i\sum_s p_s \cdot x_s}\,\Delta(x_i)\,,
    \qquad
    \mathcal{B}(p_i)
    := \int \prod_{r=1}^4 d^4 x_r\;
       e^{-i\sum_s p_s \cdot x_s}\,\mathcal{B}(x_i)\,.
\end{equation}
Substituting~\eqref{eq:B1_x_short_app}:
\begin{equation}
\label{eq:B1_fourier_expand_app}
    \mathcal{B}(p_i)
    = 4\,\Delta(p_i) + \sum_{r=1}^4 I_r(p_i)\,,
\end{equation}
where
$I_r(p_i)
:= \int \prod_s d^4 x_s\;
e^{-i\sum_t p_t \cdot x_t}\,
x_r^\mu \partial_{x_r^\mu} \Delta(x_i)$.
Integrating $I_r$ by parts in $x_r$ (dropping the spacetime
boundary term) and using
$x_r^\mu e^{-ip_r \cdot x_r}
= i\,\frac{\partial}{\partial p_{r\mu}} e^{-ip_r \cdot x_r}$
gives
\begin{equation}
\label{eq:Ir_result_app}
    I_r = -\left(4 + p_r^\mu
          \frac{\partial}{\partial p_r^\mu}\right)
          \Delta(p_i)\,.
\end{equation}
Substituting into~\eqref{eq:B1_fourier_expand_app}:
\begin{equation}
    \mathcal{B}(p_i)
    = -12\,\Delta(p_i)
    - \sum_{r=1}^4 p_r^\mu
      \frac{\partial}{\partial p_r^\mu} \Delta(p_i)\,.
\end{equation}
This is the Fourier transform of the unamputated position space
dilatation. To pass to the amputated quantity, we strip off the four
external propagators and the overall momentum conserving delta
function. Each massless external leg contributes scaling degree $-2$
in momentum, so amputation contributes $+2$ per leg, or $+8$ total.
The momentum conserving delta function $\delta^4(\sum_r p_r)$
contributes scaling degree $-4$. Together these account for the
residual constant: $-12 + 8 + 4 = 0$. The amputated result is
therefore
\begin{equation}
    \mathcal{B}^{\text{c.a.}}(p_i)
    = -\sum_{r=1}^4 p_r^\mu
      \frac{\partial}{\partial p_r^\mu} \Delta(p_i)\,,
\end{equation}
which is~\eqref{eq:B1_target_app}.

\paragraph{Remark.}
The identity~\eqref{eq:B1_target_app} receives corrections of
order $\mu^2/\Lambda^2$ from the compact support of the NN-FT
propagator, which are negligible in the regime
$|p_r| \ll \Lambda$ used to extract the beta function.

\subsection{One-loop integral at the symmetric point}
\label{app:loop_integral}

We derive the amputated one-loop vertex $\Delta^{(2)}(\mu)$ from
Wick contractions and evaluate the resulting loop integral.

\paragraph{From Wick contractions to the loop integral.}
The connected one-loop four-point function arises from the
expectation $\langle X \rangle_0^{\text{conn}}$ with
$X = \frac{1}{2}V_4^2\,\mathcal{O}_4$. In the $s$-channel,
external fields $\phi(x_1)$ and $\phi(x_2)$ contract with one
vertex at position $z$, and $\phi(x_3)$ and $\phi(x_4)$ contract
with the other vertex at position $w$, with two internal lines
connecting $z$ and $w$. The unamputated position space contribution
is
\begin{equation}
    \lambda^2 \int d^4z\,d^4w\;
    G_0(z{-}x_1)\,G_0(z{-}x_2)\,G_0(z{-}w)^2\,
    G_0(w{-}x_3)\,G_0(w{-}x_4)\,,
\end{equation}
where $G_0$ is the free NN-FT propagator. Fourier transforming in
the external coordinates $x_r \to p_r$ turns each external leg into
$G_0(p_r)\,e^{-ip_r \cdot z}$ (or $w$ for legs $3, 4$). Dividing
out the four external propagators (amputation) and the overall
momentum conserving delta function leaves
\begin{equation}
    \lambda^2 \int d^4z\,d^4w\;G_0(z{-}w)^2\,
    e^{-i(p_1+p_2)\cdot z}\,e^{-i(p_3+p_4)\cdot w}\,.
\end{equation}
Fourier expanding the two internal propagators,
\begin{equation}
    G_0(z{-}w)^2
    = \int_{|k_1|,\,|k_2| < \Lambda}
      \frac{d^4k_1}{(2\pi)^4}\frac{d^4k_2}{(2\pi)^4}\;
      \frac{e^{i(k_1+k_2)\cdot(z-w)}}{k_1^2\,k_2^2}\,.
\end{equation}
The $z$-integral produces $\delta^4(k_1 + k_2 - q)$ with
$q = p_1 + p_2$, fixing $k_2 = q - k_1$; the $w$-integral produces
$\delta^4(\sum_r p_r)$, enforcing overall momentum conservation.
Integrating out $k_2$ and renaming $k_1 \to k$:
\begin{equation}
    I_s(\mu) = \int_{|k| < \Lambda}
    \frac{d^4k}{(2\pi)^4}\;
    \frac{1}{k^2(k - q)^2}\,,
\end{equation}
where $q^2 = (p_1 + p_2)^2$. Strictly, the second propagator also
carries a cutoff $|k - q| < \Lambda$; in the regime
$\mu \ll \Lambda$ this constraint only affects momenta near the
boundary and contributes to the $O(\mu^2/\Lambda^2)$ corrections.

The $t$ and $u$ channels give the same integral at the Euclidean
symmetric point, where $s = t = u = 4\mu^2/3$. Summing the three
channels with the combinatorial weight from the Wick pairings:
\begin{equation}
    \Delta^{(2)}(\mu) = \frac{3\lambda^2}{2}\,I(\mu)\,,
    \qquad
    I(\mu) := I_s(\mu)\big|_{q^2 = 4\mu^2/3}\,.
\end{equation}

\paragraph{Evaluation of the loop integral.}
We evaluate $I(\mu)$ using a Feynman parameter. Write
\begin{equation}
    \frac{1}{k^2(k-q)^2}
    = \int_0^1 dx\;
      \frac{1}{\bigl[k^2(1-x) + (k-q)^2 x\bigr]^2}
    = \int_0^1 dx\;
      \frac{1}{\bigl[(k - xq)^2 + x(1-x)q^2\bigr]^2}\,.
\end{equation}
Shifting $k \to k + xq$ and performing the $d^4k$ integral in
Euclidean space using the standard formula
\begin{equation}
    \int \frac{d^4k}{(2\pi)^4}\;
    \frac{1}{(k^2 + \Delta)^2}
    = \frac{1}{16\pi^2}\,\log\frac{\Lambda^2}{\Delta}
    + \text{finite}\,,
\end{equation}
where $\Delta = x(1-x)q^2$ and the logarithmic divergence comes
from the upper limit $|k| = \Lambda$, we obtain
\begin{equation}
    I(\mu) = \frac{1}{16\pi^2}
    \int_0^1 dx\;\log\frac{\Lambda^2}{x(1-x)q^2}
    + \text{finite}\,.
\end{equation}
The $x$-integral of the logarithm separates into
\begin{equation}
    \int_0^1 dx\;\log\frac{\Lambda^2}{x(1-x)q^2}
    = \log\frac{\Lambda^2}{q^2}
    - \int_0^1 dx\;\log\bigl[x(1-x)\bigr]\,.
\end{equation}
The second integral is a pure number:
$-\int_0^1 dx\;\log[x(1-x)] = 2$,
which is absorbed into the finite part. Setting
$q^2 = 4\mu^2/3$:
\begin{equation}
    I(\mu) = \frac{1}{16\pi^2}\,
    \log\frac{\Lambda^2}{\mu^2} + \text{finite}\,,
\end{equation}
where ``finite'' includes the $\log(4/3)$ from $q^2/\mu^2$ and the
constant from the Feynman parameter integral. Only the logarithmic
term contributes to the beta function, since
$\mu\,dI/d\mu = -1/(8\pi^2)$.

\section{Weyl anomaly: computational details}

\subsection{Scalar Propagator in Stereographic Coordinates}
\label{app:sphere_two_pt}

We confirm that
\begin{equation}
    f(x):=\sum_{\ell=1}^\infty\frac{2\ell+1}{\ell(\ell+1)}P_\ell(x)=-\log\Big(\frac{1-x}{2}\Big)+\text{const.}\;,
\end{equation}
where $x\in[-1,1]$. The Legendre differential equation gives
\begin{equation}
    \frac{d}{dx}\Big[(1-x^2)P_\ell'(x)\Big]=-\ell(\ell+1)P_\ell(x)\;,
\end{equation}
so, applying the operator $\frac{d}{dx}[(1-x^2)\frac{d}{dx}]$ to $f$ term by term,
\begin{equation}
    \frac{d}{dx}\Big[(1-x^2)f'(x)\Big]=-\sum_{\ell=1}^\infty (2\ell+1)P_\ell(x)\;.\label{eq:legendre1}
\end{equation}
The completeness relation for Legendre polynomials,
\begin{equation}
    \delta(x-x')=\sum_{\ell=0}^\infty \frac{2\ell+1}{2}P_\ell(x) P_\ell(x')\;,\label{eq:delta_identity}
\end{equation}
follows from the orthogonality relation $\int_{-1}^1 dx\;P_\ell(x) P_{\ell'}(x)=\frac{2}{2\ell+1}\delta_{\ell,\ell'}$ and completeness of the Legendre polynomials in $L^2([-1,1])$. Setting $x'=1$ and using $P_\ell(1)=1$ and $P_0(x)=1$, the right-hand side of~\eqref{eq:legendre1} becomes $1-2\delta(x-1)$. Away from $x=1$, this gives the ODE
\begin{equation}
    \frac{d}{dx}\Big[(1-x^2)f'(x)\Big]=1\;,
\end{equation}
which integrates to $(1-x^2)f'(x)=x+C_1$. Since $f$ is regular at $x=-1$ (the series converges there by the alternating series test), the left-hand side vanishes at $x=-1$, fixing $C_1=1$. Then
\begin{equation}
    f'(x)=\frac{1}{1-x}\;\Longrightarrow\; f(x)=-\log\Big(\frac{1-x}{2}\Big)+\text{const.}\;,
\end{equation}
as desired.

\subsection{Ghost propagator on $S^2$}
\label{app:ghost_prop}

We verify that the ghost architecture~\eqref{eq:bc-sphere}
with the density~\eqref{eq:ghost-density} reproduces the
standard ghost OPE
$\langle b_{zz}(z)\, c^w(w)\rangle \sim 1/(z-w)$
in stereographic coordinates,
and determine the exact nonzero mode propagator on $S^2$.

\subsubsection{From modes to Green equation}

In stereographic coordinates $z = \cot(\theta/2)\,e^{i\phi}$,
the round metric is $ds^2 = 2P^{-2}\,|dz|^2$ with
$P = (1+|z|^2)/\sqrt{2}$.
The coordinate components of the ghost fields are related
to the abstract spin-weighted fields by the dyad:
\begin{equation}
b_{zz} = P^{-2}\, b\,,\qquad
c^z = P\, c\,.
\end{equation}
Restricting to the nonzero mode sector $\ell \geq 2$
(the three CKV modes $\chi_{1,m}$ are treated separately),
the density~\eqref{eq:ghost-density} gives
$\langle \beta_{\ell m}\, \chi_{\ell'm'}\rangle
= \frac{4\pi}{\mu_\ell}\,\delta_{\ell\ell'}\delta_{mm'}$,
so the coordinate component propagator is
\begin{equation}\label{eq:Gperp-modesum}
G_\perp(z,w)
:= \langle b_{zz}(z)\, c_\perp^w(w) \rangle
= \frac{4\pi\, P(w)}{P(z)^2}
\sum_{\ell=2}^{\infty}\sum_{m=-\ell}^{\ell}
\frac{{}_{-2}Y_{\ell m}^*(z)\;{}_{-1}Y_{\ell m}(w)}{\mu_\ell}\,.
\end{equation}

We now show that $G_\perp$ satisfies the standard ghost
Green equation.
The spin-lowering eigenvalue
relation~\cite{Goldberg1967}
$\bar\eth\,{}_{-1}Y_{\ell m} = -\mu_\ell\,{}_{-2}Y_{\ell m}$
and the completeness of spin-weight-$(-2)$ harmonics
(which exist only for $\ell \geq 2$, so no modes are missing)
imply
\begin{equation}\label{eq:K-green-app}
\sum_{\ell=2}^{\infty}\sum_{m=-\ell}^{\ell}
{}_{-2}Y_{\ell m}^*(z)\;{}_{-2}Y_{\ell m}(w)
= \delta(\cos\theta - \cos\theta')\,\delta(\phi - \phi')\,,
\end{equation}
from which the abstract field mode sum
$K_0 = \sum \mu_\ell^{-1}\,{}_{-2}Y^*\,{}_{-1}Y$
satisfies
$\bar\eth_w K_0 = -\delta_{(-2)}$,
where $\delta_{(-2)}$ denotes the right-hand side
of~\eqref{eq:K-green-app}.
For spin-weight $s = -1$, the
operator $\bar\eth$ takes the
form~\cite{NewmanPenrose1966,Goldberg1967}
$\bar\eth\,\eta = 2\,\partial_{\bar w}(P\,\eta)$,
so
\begin{equation}
2\,\partial_{\bar w}\bigl(P(w)\, K_0\bigr)
= -\delta_{(-2)}
= -\frac{P_0(w)^2}{2}\,\delta^{(2)}(z - w)\,,
\end{equation}
where $P_0 = 1 + |z|^2 = \sqrt{2}\,P$ and
$\delta^{(2)}$ is the flat delta with
$\int d^2z\,\delta^{(2)} = 1$
(with $d^2z = 2\,dx\,dy$).
From~\eqref{eq:Gperp-modesum},
$P(w)\,K_0 = P(z)^2\, G_\perp / (4\pi)$,
and substituting:
\begin{equation}
\frac{P(z)^2}{2\pi}\,\partial_{\bar w} G_\perp
= \frac{P_0(w)^2}{2}\,\delta^{(2)}(z-w)
= P(w)^2\,\delta^{(2)}(z-w)\,.
\end{equation}
Since $P(w) = P(z)$ on the support of the delta function,
\begin{equation}\label{eq:green-Gperp}
\boxed{\partial_{\bar w}\, G_\perp(z,w)
= 2\pi\,\delta^{(2)}(z - w)\,.}
\end{equation}
This is the defining equation of the ghost propagator
for the action $S = \frac{1}{2\pi}\int d^2z\;b_{zz}\,\bar\partial\, c^z$.

\subsubsection{CKV projection and exact propagator}

By~\eqref{eq:green-Gperp}, any other local right
inverse of $\bar\partial$ differs from $G_\perp$
by a function holomorphic in $w$.
Since $\partial_{\bar w}[1/(z-w)] = 2\pi\,\delta^{(2)}(z-w)$,
the most general form is
\begin{equation}\label{eq:Gperp-ansatz}
G_\perp(z,w) = \frac{1}{z - w}
+ A_0(z) + A_1(z)\,w + A_2(z)\,w^2\,,
\end{equation}
where $\{1, w, w^2\}$ span the conformal Killing
vectors of $S^2$ in stereographic coordinates.
(Higher polynomials in $w$ are excluded because they
would produce a non-normalizable spin-weight-$(-1)$
section on the sphere.)
The mode sum~\eqref{eq:Gperp-modesum}
projects out the three $\ell = 1$ modes by construction,
so $G_\perp$ is orthogonal to each CKV:
\begin{equation}\label{eq:ortho-cond}
\int_{\mathbb{C}} \frac{d^2w}{(1+|w|^2)^4}\;
\bar w^{\,n}\, G_\perp(z,w) = 0\,,
\qquad n = 0, 1, 2\,,
\end{equation}
where the measure
$d^2w/(1+|w|^2)^4 = d\Omega\, P(w)^{-2}$
is the $L^2$ pairing for spin-weight-$(-1)$ sections.

The coefficients $A_n$ are determined by
substituting~\eqref{eq:Gperp-ansatz}
into~\eqref{eq:ortho-cond}.
By rotational symmetry,
$\int d^2w\,(1+|w|^2)^{-4}\,\bar w^n\, w^k = 0$
for $k \neq n$, so each $A_n$ is fixed independently:
\begin{equation}
A_n(z) = -\frac{\displaystyle
\int \frac{d^2w}{(1+|w|^2)^4}\;
\frac{\bar w^{\,n}}{z - w}}
{\displaystyle
\int \frac{d^2w}{(1+|w|^2)^4}\;|w|^{2n}}\,.
\end{equation}
Writing $w = r\,e^{i\varphi}$, the angular integral
evaluates to
\begin{equation}
\int_0^{2\pi}
\frac{e^{-in\varphi}\,d\varphi}{z - r\,e^{i\varphi}}
= \begin{cases}
\displaystyle \frac{2\pi\, r^n}{z^{n+1}}\,, & r < |z|\,,\\
0\,, & r > |z|\,,
\end{cases}
\end{equation}
for $n = 0, 1, 2$, reducing each $A_n$ to a ratio of
explicit radial integrals of the form
$\int_0^{\rho} r^{2n+1}\,dr/(1+r^2)^4$.
Evaluating these (the integrands are rational and
integrate to elementary functions), we obtain
\begin{equation}
A_0 = -\frac{\bar z\,(|z|^4+3|z|^2+3)}{(1+|z|^2)^3}\,,
\quad
A_1 = -\frac{\bar z^2\,(|z|^2+3)}{(1+|z|^2)^3}\,,
\quad
A_2 = -\frac{\bar z^3}{(1+|z|^2)^3}\,.
\end{equation}
Substituting into~\eqref{eq:Gperp-ansatz} and
factoring:
\begin{equation}\label{eq:Gperp-exact}
\boxed{G_\perp(z,w)
= \frac{(1 + \bar z\, w)^3}{(1+|z|^2)^3}\;
\frac{1}{z - w}\,.}
\end{equation}
At short distance the prefactor
$(1+\bar z w)^3/(1+|z|^2)^3 \to 1$
as $w \to z$, confirming the OPE
$\langle b_{zz}(z)\, c_\perp^w(w)\rangle
= \frac{1}{z-w} + O\!\bigl((z-w)^0\bigr)$.
After the standard treatment of the three
$c$-ghost zero modes on $S^2$
(saturated by operator insertions), the local
Wick contraction is the familiar
$b_{zz}(z)\,c^w(w) \sim 1/(z-w)$.

For the antiholomorphic sector, the identical
analysis gives
$\langle \bar b_{\bar z\bar z}(z)\,
\bar c_\perp^{\bar w}(w)\rangle
= \frac{(1 + z\bar w)^3}{(1+|z|^2)^3}\,
\frac{1}{\bar z - \bar w}$,
and all mixed chirality correlators vanish.

\subsection{Weyl scaling of spin-weight-$s$ sections}
\label{app:weyl_scaling}

We prove that the $L^2$ inner product on
a tensor field with $t$ net lower indices on a
two-dimensional Riemannian manifold $(M, g)$ scales as
$e^{(2-2t)\epsilon}$ under a constant Weyl rescaling
$g \to e^{2\epsilon}g$, where
$t = n_{\text{lower}} - n_{\text{upper}}$
counts lower minus upper indices regardless of
holomorphic or antiholomorphic type.

Let $(M, g)$ be a compact two-dimensional Riemannian
manifold.
In local complex coordinates $(z, \bar z)$,
the metric takes the form
$ds^2 = 2g_{z\bar z}\, dz\, d\bar z$.
A field of spin-weight $s$ on $M$ is a section of
a complex line bundle whose local representative
$\eta$ carries a definite index structure.
For the fields appearing in the bosonic string:
\begin{center}
\begin{tabular}{cccc}
Field & Representative & $s$ & $t$ \\
\hline
$X^\mu$ & $\phi$ (scalar) & $0$ & $0$ \\
$b$ & $b_{zz}$ (quad.\ diff.) & $+2$ & $+2$ \\
$c$ & $c^z$ (vector) & $-1$ & $-1$ \\
$\bar b$ & $\bar b_{\bar z\bar z}$ (antihol.\ quad.\ diff.) & $-2$ & $+2$ \\
$\bar c$ & $\bar c^{\bar z}$ (antihol.\ vector) & $+1$ & $-1$
\end{tabular}
\end{center}
For holomorphic sections $t = s$; for antiholomorphic
sections $t = -s$.
Both chiralities of each ghost carry the same value of $t$.

The inner product on a field with $t \geq 0$
(net lower indices) is built from the area element
$\sqrt{g}\, d^2 x$ and $t$ contractions with the
inverse metric:
\begin{equation}\label{eq:norm-lower}
\langle \eta, \eta \rangle
= \int_M \sqrt{g}\, d^2 x\;
(g^{z\bar z})^t\,
|\eta_{\underbrace{z\cdots z}_{t}}|^2\,.
\end{equation}
For $t < 0$ (net upper indices), the $|t|$
upper index pairs are contracted with the metric:
\begin{equation}\label{eq:norm-upper}
\langle \eta, \eta \rangle
= \int_M \sqrt{g}\, d^2 x\;
(g_{z\bar z})^{|t|}\,
|\eta^{\overbrace{z\cdots z}^{|t|}}|^2\,.
\end{equation}
These formulas apply whether the indices are holomorphic
or antiholomorphic: $g^{z\bar z}$ contracts a pair of
lower indices of either type, and $g_{z\bar z}$ contracts
a pair of upper indices of either type.

Under $g \to e^{2\epsilon}g$, the building blocks
transform as
\begin{equation}
\sqrt{g} \to e^{2\epsilon}\sqrt{g}\,,
\qquad
g^{z\bar z} \to e^{-2\epsilon} g^{z\bar z}\,,
\qquad
g_{z\bar z} \to e^{+2\epsilon} g_{z\bar z}\,,
\end{equation}
where the power of $2$ in the area element is
specific to two dimensions.
For $t \geq 0$, the inner product~\eqref{eq:norm-lower}
scales as
\begin{equation}
\langle \eta, \eta \rangle
\;\to\;
e^{2\epsilon}\cdot (e^{-2\epsilon})^t\,
\langle \eta, \eta \rangle
= e^{(2-2t)\epsilon}\,
\langle \eta, \eta \rangle\,.
\end{equation}
For $t < 0$, the inner product~\eqref{eq:norm-upper}
scales as
\begin{equation}
\langle \eta, \eta \rangle
\;\to\;
e^{2\epsilon}\cdot (e^{+2\epsilon})^{|t|}\,
\langle \eta, \eta \rangle
= e^{(2+2|t|)\epsilon}\,
\langle \eta, \eta \rangle
= e^{(2-2t)\epsilon}\,
\langle \eta, \eta \rangle\,,
\end{equation}
using $|t| = -t$ for $t < 0$.
In both cases:
\begin{equation}\label{eq:weyl-scaling-result}
\boxed{\langle \eta, \eta \rangle
\;\to\; e^{(2-2t)\epsilon}\,
\langle \eta, \eta \rangle\,.}
\end{equation}

\subsubsection{Consequence for orthonormal bases}

Let $\{f_n\}$ be an orthonormal basis
with respect to $g$,
so $\langle f_n, f_{n'}\rangle = \delta_{nn'}$.
Under Weyl rescaling, unit normalization requires
\begin{equation}
\widetilde{f}_n = e^{-(1-t)\epsilon}\, f_n\,.
\end{equation}
A field expanded as
$\eta = \sum_n c_n\, f_n$ on the original background
becomes
$\eta = \sum_n c_n\, \widetilde{f}_n
= \sum_n [e^{-(1-t)\epsilon}\, c_n]\, f_n$
on the rescaled background.
The absorption condition
$\eta_{\theta}^{\text{new}} = \eta_{\theta'}^{\text{old}}$
then requires the coefficients to transform as
\begin{equation}\label{eq:xi-from-weyl}
c_n' = e^{-(1-t)\epsilon}\, c_n
\quad\Longrightarrow\quad
\xi^{c_n} = -(1-t)\, c_n\,,
\end{equation}
which is used above equation (\ref{eq:weyl-generators}) in the
main text.
As a check:
$t = 0$ gives $\xi = -c_n$ (scalar);
$t = +2$ gives $\xi = +c_n$ ($b$ and $\bar b$);
$t = -1$ gives $\xi = -2c_n$ ($c$ and $\bar c$).

\subsection{Evaluation of spectral zeta functions}
\label{app:zeta}

We evaluate the spectral zeta functions
$\zeta_X(0)$ and $\zeta_{bc}(0)$ defined in
Section~\ref{sec:weyl}.
The method is the same in both cases: a change
of variables reduces the sum to Hurwitz zeta
functions, whose values at negative integers
are given by Bernoulli polynomials.

The Hurwitz zeta function is
$\zeta_H(s, a) = \sum_{n=0}^\infty (n+a)^{-s}$,
defined by analytic continuation from
$\text{Re}(s) > 1$.
At negative integers it satisfies~\cite{Apostol1976}
\begin{equation}\label{eq:hurwitz-bernoulli}
\zeta_H(-n, a) = -\frac{B_{n+1}(a)}{n+1}\,,
\end{equation}
where $B_n(x)$ is the $n$-th Bernoulli polynomial.
In particular,
$\zeta_H(-1, a) = -B_2(a)/2$
with $B_2(x) = x^2 - x + \frac{1}{6}$.
Near $s = 0$, $\zeta_H$ has the Laurent expansion
$\zeta_H(1 + \epsilon, a) = 1/\epsilon + O(1)$.

\subsubsection{Scalar sector}

The scalar zeta function is
$\zeta_X(s) = \sum_{\ell=1}^\infty
(2\ell+1)\,[\ell(\ell{+}1)]^{-s}$.
Writing $2\ell+1 = 2(\ell+\frac{1}{2})$
and $\ell(\ell{+}1) = (\ell+\frac{1}{2})^2 - \frac{1}{4}$,
then factoring:
\begin{equation}
\zeta_X(s)
= 2\sum_{\ell=1}^\infty
\bigl(\ell + \tfrac{1}{2}\bigr)^{1-2s}
\Bigl(1 - \frac{1}{4(\ell+\frac{1}{2})^2}\Bigr)^{\!-s}.
\end{equation}
Expanding $(1-x)^{-s} = 1 + sx + O(x^2)$
with $x = 1/[4(\ell+\frac{1}{2})^2]$:
\begin{equation}
\zeta_X(s)
= 2\biggl[
\zeta_H\bigl(2s{-}1, \tfrac{3}{2}\bigr)
+ \frac{s}{4}\,
\zeta_H\bigl(2s{+}1, \tfrac{3}{2}\bigr)
+ O(s^2)
\biggr],
\end{equation}
where we have used
$\sum_{\ell=1}^\infty (\ell+\frac{1}{2})^{-s}
= \zeta_H(s, \frac{3}{2})$. At $s = 0$, the first term gives
$2\,\zeta_H(-1, \frac{3}{2})
= -B_2(\frac{3}{2})$.
For the second term,
$\zeta_H(1 + 2s, \frac{3}{2})$ has a simple pole
at $s = 0$ with Laurent expansion
$\frac{1}{2s} + O(1)$, so
$\frac{s}{4} \cdot \frac{1}{2s} = \frac{1}{8}$,
contributing $2 \cdot \frac{1}{8} = \frac{1}{4}$.
The $O(s^2)$ terms do not contribute at $s = 0$.
Therefore
\begin{equation}
\zeta_X(0)
= -B_2\Big(\frac{3}{2}\Big) + \frac{1}{4}
=-\frac{2}{3}\,,
\end{equation}
where $B_2(\frac{3}{2}) = \frac{11}{12}$.

\subsubsection{Ghost sector}

The ghost zeta function is
$\zeta_{bc}(s) = \sum_{\ell=2}^\infty
(2\ell+1)\,\mu_\ell^{-s}$
with $\mu_\ell = \sqrt{(\ell{-}1)(\ell{+}2)}$.
Writing $2\ell+1 = 2(\ell+\frac{1}{2})$
and $(\ell{-}1)(\ell{+}2) = (\ell+\frac{1}{2})^2 - \frac{9}{4}$,
then factoring:
\begin{equation}
\zeta_{bc}(s)
= 2\sum_{\ell=2}^\infty
\bigl(\ell + \tfrac{1}{2}\bigr)^{1-s}
\Bigl(1 - \frac{9}{4(\ell+\frac{1}{2})^2}\Bigr)^{\!-s/2}.
\end{equation}
Expanding $(1-x)^{-s/2} = 1 + \frac{s}{2}x + O(x^2)$
with $x = 9/[4(\ell+\frac{1}{2})^2]$:
\begin{equation}
\zeta_{bc}(s)
= 2\biggl[
\zeta_H\bigl(s{-}1, \tfrac{5}{2}\bigr)
+ \frac{9s}{8}\,
\zeta_H\bigl(s{+}1, \tfrac{5}{2}\bigr)
+ O(s^2)
\biggr]\;.
\end{equation}
At $s = 0$, the first term gives
$2\,\zeta_H(-1, \frac{5}{2})
= -B_2(\frac{5}{2})$.
For the second term,
$\zeta_H(1+s, \frac{5}{2}) = \frac{1}{s} + O(1)$
near $s = 0$, so
$\frac{9s}{8}\cdot\frac{1}{s} = \frac{9}{8}$,
contributing $2\cdot\frac{9}{8} = \frac{9}{4}$.
Therefore
\begin{equation}
\zeta_{bc}(0)
= -B_2\bigl(\tfrac{5}{2}\bigr) + \frac{9}{4}
= -\frac{5}{3}\,,
\end{equation}
where $B_2(\frac{5}{2})
= \frac{47}{12}$.

\section{Berezinskii--Kosterlitz--Thouless (BKT) example}\label{app:bkt}
As we noted in Section~\ref{sec:higher-form}, winding symmetry can be approached by considering the $(d-2)$-form which is dual to the compact scalar $\chi$.
In general dimension, we denote this dual field by $\widetilde B$, and the charged operator is schematically
\begin{equation}
V_m(\Sigma) \sim \exp\!\left(i m\int_\Sigma \widetilde B\right) \,.
\end{equation}
This strategy simplifies in $d = 2$, where the winding symmetry is an ordinary zero-form symmetry, and vortex defects are local operators.
This is the setting relevant for the Berezinskii--Kosterlitz--Thouless (BKT) transition, which is a phase transition in the $2d$ XY model, whose continuum version flows in the IR to the theory of a $2d$ scalar.
This transition was realized via a NN-FT presentation in~\cite{Ferko:2026ken}.
For this example, in the dual description one has
\begin{equation}
V_m(x)\sim e^{im\widetilde\chi(x)}\,,
\end{equation}
where $\widetilde\chi$ is the dual compact scalar.  

The NN-FT construction of the BKT transition~\cite{Ferko:2026ken} fits precisely into the
mixed form~\eqref{eq:hf-mixed-expectation}.
The low-temperature behavior of this model is controlled by a spin-wave action
\begin{equation}
S_{\text{sw}} = \frac{K_R}{2} \int d^2 x \, \left( \nabla \theta_{\text{sw}} \right)^2 \,,
\end{equation}
where $K_R$ is a ``dimensionless stiffness.''\footnote{
Although we use the symbol $\theta$ for the compact field in the spin-wave model for historical reasons, we trust that the reader will not confuse this symbol with the NN-FT parameters that are also labeled $\theta$ in this work.}
In order to describe the BKT transition, which is driven by unbinding of vortices (non-trivial winding configurations) at high temperatures, the spin-wave component of the architecture is supplemented with an explicit vortex term as $\theta(x)=b\,\theta_{\text{sw}}(x)+\theta_v(x)$.
The combined field $\theta ( x )$ is a compact boson obeying $\theta ( x ) \sim \theta ( x ) + 2 \pi$.
While the spin-wave architecture is defined by a random Fourier feature field of the type introduced in~\cite{Halverson:2021aot}, the vortex term includes additional parameters
\begin{equation}
Q=\{(x_a,m_a)\}_{a=1}^{N_v}\,, \qquad m_a=\pm1\,.
\end{equation}

The corresponding NN-FT expectation value has the schematic form
\begin{equation}
\langle\mathcal O\rangle = \sum_{N_+,N_-\ge 0} \frac{y^{N_++N_-}}{N_+!\,N_-!} \int d\theta_{\rm sw}\,P_{\rm sw}(\theta_{\rm sw}) \prod_{a=1}^{N_v} d^2x_a\; e^{-S_{\rm vort}(Q)} \, \mathcal O[\theta_{\theta_{\rm sw},Q}] \,,
\end{equation}
with $N_v=N_++N_-$, charges $m_a=\pm1$, and the neutrality condition
\begin{equation}
\sum_a m_a=0
\end{equation}
on compact space.
The smooth spin-wave parameters obey the ordinary Schwinger--Dyson identities with their own breaking function, while the winding Ward identity is encoded in the sector observable
\begin{equation}
W_Q(\gamma)=\sum_a m_a\,\operatorname{Wind}(\gamma,x_a) = \sum_{x_a\in \text{int}(\gamma)}m_a \,.
\end{equation}
A neutral vortex--antivortex pair can be separated by $\gamma$, so $U_\alpha(\gamma)[Q]\neq1$ for generic $\alpha$.

For $y=0$, the vacuum ensemble is supported on source free sectors and the continuous winding symmetry is exact.
For $y\neq0$, charged vortex sectors are included in the vacuum measure.
A contractible loop can then enclose dynamical vortex charge, so the continuous winding symmetry is explicitly broken.
In the dual sine-Gordon description this is the statement that
\begin{equation}
2y\int d^2x\,\cos\widetilde\chi
\end{equation}
is not invariant under the continuous shift $\widetilde\chi\mapsto\widetilde\chi+\alpha$, except for the residual subgroup allowed by the vortex charge lattice $\mathbb Z_k$.

Thus the BKT transition is not a failure of the linking Ward identity for external vortex operators.
Rather, it is a phase transition between (i) a regime where the charged vortex fugacity is irrelevant and the continuous winding symmetry re-emerges at the infrared fixed line, and (ii) a regime where the charged vortex deformation is relevant and disorders the compact boson.
Again, the Schwinger--Dyson identities continue to hold within each continuous sector, but the topological physics is controlled by the sector measure over discrete labels.

\section{Enhanced $SU(2)$ symmetry from the NN-FT Ward kernel}\label{sec:su2}

The $SU(2)_L \times SU(2)_R$ enhancement at the self-dual radius can be phrased directly in the breaking function language.
The key point is that the additional currents appear precisely when certain NN-FT Ward-breaking insertions enter the kernel of the parameter space Ward operator.

It is useful to introduce the dimensionless chiral charges
\begin{equation}
q_L^{(R)}(n,w):=\sqrt{\alpha'}\,p_L^{(R)}(n,w) =\sqrt{\alpha'}\,\frac{n}{R}+\frac{R}{\sqrt{\alpha'}}\,w \,,
\qquad q_R^{(R)}(n,w):=\sqrt{\alpha'}\,p_R^{(R)}(n,w) =\sqrt{\alpha'}\,\frac{n}{R}-\frac{R}{\sqrt{\alpha'}}\,w \,.
\label{eq:su2-qLR}
\end{equation}
The conformal weights of $V_{n,w}^{(R)}$ are
\begin{equation}
h_{n,w}^{(R)}=\frac14\big(q_L^{(R)}(n,w)\big)^2 \,,
\qquad \bar h_{n,w}^{(R)}=\frac14\big(q_R^{(R)}(n,w)\big)^2 \,.
\end{equation}
At generic radius the smooth zero mode shifts give only the Cartan symmetry $U(1)_L\times U(1)_R$.
Writing
\begin{equation}
x_L=\frac12(x_0+\widetilde x_0) \,,
\qquad x_R=\frac12(x_0-\widetilde x_0) \,,
\end{equation}
the infinitesimal chiral shifts are generated by
\begin{equation}
\Xi_0^L=\sqrt{\alpha'}\,(\partial_{x_0}+\partial_{\widetilde x_0}) \,,
\qquad \Xi_0^R=\sqrt{\alpha'}\,(\partial_{x_0}-\partial_{\widetilde x_0}) \,.
\end{equation}
Because the zero mode density in~\eqref{eq:tdual-hf-mixed-ensemble} is uniform, both breaking functions vanish:
\begin{equation}
B_{\Xi_0^L}=B_{\Xi_0^R}=0 \,.
\end{equation}
Acting on a vertex operator,
\begin{equation}
\delta_{\Xi_0^L}V_{n,w}^{(R)} =i\,q_L^{(R)}(n,w)\,V_{n,w}^{(R)} \,,
\qquad \delta_{\Xi_0^R}V_{n,w}^{(R)} =i\,q_R^{(R)}(n,w)\,V_{n,w}^{(R)} \,,
\end{equation}
which gives the generic compact boson charge selection rules.

The non-Cartan enhancement is detected by a different Ward kernel.
Consider the oscillator part of the chiral finite mode architecture~\eqref{eq:compact_chiral_features}. Worldsheet translations of the left- and right-moving oscillator networks are absorbed by the parameter space vector fields
\begin{equation}
\Xi_L=\sum_{m=1}^{M}m \left( b_m^L\frac{\partial}{\partial a_m^L} -a_m^L\frac{\partial}{\partial b_m^L}
\right) \,,
\qquad \Xi_R=\sum_{m=1}^{M}m \left( b_m^R\frac{\partial}{\partial a_m^R} -a_m^R\frac{\partial}{\partial b_m^R} \right) \,.
\label{eq:su2-chiral-translation-vectors}
\end{equation}
Indeed,
\begin{equation}
\delta_{\Xi_L}X_L^{\rm osc}(u)=\partial_u X_L^{\rm osc}(u) \,,
\qquad \delta_{\Xi_R}X_R^{\rm osc}(v)=\partial_v X_R^{\rm osc}(v) \,.
\end{equation}
The Gaussian oscillator density is rotationally invariant in each $(a_m,b_m)$-plane and the vector fields~\eqref{eq:su2-chiral-translation-vectors} are divergence free, hence
\begin{equation}
B_{\Xi_L}=B_{\Xi_R}=0 \,.
\label{eq:su2-osc-breaking-zero}
\end{equation}
If one keeps the classical zero mode slopes in~\eqref{eq:lrmovers}, $\Xi_L$ and $\Xi_R$ are augmented by the corresponding constant translations of $x_L$ and $x_R$; the zero mode density is still uniform, so~\eqref{eq:su2-osc-breaking-zero} is unchanged.

Now take a candidate vertex current
\begin{equation}
J_{\ell}^{(R)}(u,v) = \ :\!\exp\!\left(i p_L^{(R)}(\ell)X_L(u) +i p_R^{(R)}(\ell)X_R(v) \right)\!: \,,
\qquad \ell=(n,w) \,.
\end{equation}
A left-moving current must be invisible to the right-moving Ward operator.
Since $B_{\Xi_R}=0$, the relevant NN-FT breaking insertion is the part of the contracted Schwinger--Dyson identity in which $\Xi_R$ acts on the candidate current:
\begin{equation}
\mathcal W_{L,\ell}^{\rm NN}[\mathcal O] := \left\langle \delta_{\Xi_R}J_{\ell}^{(R)}\,\mathcal O \right\rangle_R = i\,p_R^{(R)}(\ell) \left\langle :\!\partial_v X_R\,J_{\ell}^{(R)}\!:\,\mathcal O \right\rangle_R \,.
\label{eq:su2-left-current-breaking}
\end{equation}
Similarly, a right-moving current is tested by
\begin{equation}
\mathcal W_{R,\ell}^{\rm NN}[\mathcal O] := \left\langle \delta_{\Xi_L}J_{\ell}^{(R)}\,\mathcal O \right\rangle_R = i\,p_L^{(R)}(\ell) \left\langle :\!\partial_u X_L\,J_{\ell}^{(R)}\!:\,\mathcal O \right\rangle_R \,.
\label{eq:su2-right-current-breaking}
\end{equation}
Equations~\eqref{eq:su2-left-current-breaking} and~\eqref{eq:su2-right-current-breaking} are the NN-FT version of the current conservation test: the obstruction is computed by acting with a parameter space Ward vector field on a network observable.
They are not postulated as field space equations.

The candidate left roots are the sector labels
\begin{equation}
\ell_L^\pm=(\pm1,\pm1) \,.
\end{equation}
For these,
\begin{equation}
p_R^{(R)}(\ell_L^\pm) = \pm\left(\frac1R-\frac{R}{\alpha'}\right) \,.
\end{equation}
Thus,
\begin{equation}
\mathcal W_{L,\ell_L^\pm}^{\rm NN}=0 \quad\Longleftrightarrow\quad R=\sqrt{\alpha'} \,.
\end{equation}
At this same radius,
\begin{equation}
p_L^{(R)}(\ell_L^\pm)=\pm\frac{2}{\sqrt{\alpha'}} \,,
\qquad
(h,\bar h)=(1,0) \,,
\end{equation}
so the two operators become genuine dimension one left-moving currents,
\begin{equation}
J_L^\pm(u) = \ :\!\exp\!\left(\pm\frac{2i}{\sqrt{\alpha'}}X_L(u)\right)\!: \,.
\end{equation}
Likewise, the candidate right roots
\begin{equation}
\ell_R^\pm=(\pm1,\mp1)
\end{equation}
obey
\begin{equation}
p_L^{(R)}(\ell_R^\pm) = \pm\left(\frac1R-\frac{R}{\alpha'}\right) \,,
\end{equation}
so their NN-FT Ward-breaking insertions vanish only at $R=\sqrt{\alpha'}$, where
\begin{equation}
J_R^\pm(v) = \ :\!\exp\!\left(\pm\frac{2i}{\sqrt{\alpha'}}X_R(v)\right)\!: \,,
\qquad (h,\bar h)=(0,1) \,.
\end{equation}

Therefore, the self-dual radius is characterized intrinsically as the point where the NN-FT Ward kernel enlarges:
\begin{equation}
\ker \mathcal W^{\rm NN}(R): \quad \{J_L^3,J_R^3\} \quad\longrightarrow\quad \{J_L^3,J_L^\pm\}\oplus\{J_R^3,J_R^\pm\} \,,
\qquad R=\sqrt{\alpha'} \,.
\label{eq:su2-kernel-enhancement}
\end{equation}
This is the breaking function statement of enhanced symmetry.
Away from the self-dual radius the candidate root operators fail the NN-FT Ward kernel test because they carry the wrong chiral charge.
At the self-dual radius the corresponding breaking insertions vanish as operator insertions in the NN-FT ensemble.

The action on vertex operators is also naturally sector theoretic.
At $R=\sqrt{\alpha'}$,
\begin{equation}
q_L=n+w \,,\qquad q_R=n-w \,.
\end{equation}
The Cartan charges act as
\begin{equation}
Q_L^3\,V_{n,w}=\frac12(n+w)V_{n,w} \,,
\qquad Q_R^3\,V_{n,w}=\frac12(n-w)V_{n,w} \,.
\end{equation}
The non-Cartan charges are contour integrals of the root currents.
Using only the Gaussian NN-FT covariance and zero mode averaging, their residue action is the finite relabeling
\begin{equation}
Q_L^\pm:\ (n,w)\mapsto(n\pm1,w\pm1) \,,
\qquad Q_R^\pm:\ (n,w)\mapsto(n\pm1,w\mp1) \,,
\label{eq:su2-sector-ladders}
\end{equation}
on the residue supported sectors.
Equivalently,
\begin{equation}
Q_L^\pm:\ q_L\mapsto q_L\pm2,\quad q_R\mapsto q_R \,,
\qquad Q_R^\pm:\ q_R\mapsto q_R\pm2,\quad q_L\mapsto q_L \,.
\end{equation}
Together with the Cartan zero mode Ward identities, these maps obey
\begin{equation}
[Q_L^3,Q_L^\pm]=\pm Q_L^\pm \,,
\qquad [Q_L^+,Q_L^-]=2Q_L^3 \,,
\end{equation}
and similarly in the right-moving sector, while all left generators commute with all right generators.
Thus the enhanced symmetry is
\begin{equation}
SU(2)_L\times SU(2)_R \,.
\end{equation}
Finally, the finite breaking functions introduced in Section~\ref{sec:T} identify the $\mathbb Z_2$ T-duality at the self-dual radius as a Weyl reflection inside this enhanced group.
The right Weyl reflection is
\begin{equation}
\mathsf r_R:\ (n,w)\mapsto(w,n) \,,
\end{equation}
which sends $q_R\mapsto-q_R$ and fixes $q_L$.
The left Weyl reflection is
\begin{equation}
\mathsf r_L:\ (n,w)\mapsto(-w,-n) \,,
\end{equation}
which sends $q_L\mapsto-q_L$ and fixes $q_R$.
Their finite breaking functions are
\begin{equation}
\mathfrak B_{\mathsf r_R}(n,w) = 1-\frac{P_R(w,n)}{P_R(n,w)} \,,
\qquad \mathfrak B_{\mathsf r_L}(n,w) = 1-\frac{P_R(-w,-n)}{P_R(n,w)} \,.
\end{equation}
Both vanish identically at $R=\sqrt{\alpha'}$.
Thus the same finite breaking function formalism which detects T-duality also detects the Weyl reflections of the enhanced $SU(2)_L\times SU(2)_R$, while the local Ward kernel test~\eqref{eq:su2-left-current-breaking}--\eqref{eq:su2-kernel-enhancement} supplies the genuinely continuous
current enhancement.

\bibliography{refs}
\end{document}